\documentclass[twocolumn]{aastex63}

\shorttitle{\textsc{CVs and AM CVns in SRG/eROSITA + Gaia}}
\shortauthors{Rodriguez et al.}
\graphicspath{{./}{figures/}}

\usepackage{float}
\usepackage{appendix}
\usepackage{xcolor}
\usepackage{amsmath}

\begin{document}

\title{Cataclysmic Variables and AM CVn Binaries in SRG/eROSITA + \textit{Gaia}: Volume Limited Samples, X-ray Luminosity Functions, and Space Densities}

\correspondingauthor{Antonio C. Rodriguez}
\email{acrodrig@caltech.edu}

\author[0000-0003-4189-9668]{Antonio C. Rodriguez}
\affiliation{Department of Astronomy, California Institute of Technology, 1200 East California Blvd, Pasadena, CA, 91125, USA}

\author[0000-0002-6871-1752]{Kareem El-Badry}
\affiliation{Department of Astronomy, California Institute of Technology, 1200 East California Blvd, Pasadena, CA, 91125, USA}

\author[0000-0003-3733-7267]{Valery Suleimanov}
\affiliation{Institut f{\"u}r Astronomie und Astrophysik, Universit{\"a}t T{\"u}bingen, Sand 1, 72076, T{\"u}bingen, Germany}

\author[0000-0001-7069-7403]{Anna F. Pala}
\affiliation{European Space Agency, European Space Astronomy Centre, Camino Bajo del Castillo s/n, Villanueva de la Cañada, E-28692 Madrid, Spain}

\author[0000-0001-5390-8563]{Shrinivas R. Kulkarni}
\affiliation{Department of Astronomy, California Institute of Technology, 1200 East California Blvd, Pasadena, CA, 91125, USA}

\author[0000-0002-2761-3005]{Boris Gaensicke}
\affiliation{Department of Physics, University of Warwick, Coventry CV4 7AL, UK}

\author[0000-0002-9709-5389]{Kaya Mori}
\affiliation{Columbia Astrophysics Laboratory, Columbia University, New York, NY, USA}

\author[0000-0003-0427-8387]{R. Michael Rich}
\affiliation{Department of Physics and Astronomy, UCLA, 430 Portola Plaza, Box 951547, Los Angeles, CA, 90095-1547, USA}

\author[0000-0002-1455-2784]{Arnab Sarkar} 
\affiliation{Institute of Astronomy, The Observatories, Madingley Road, Cambridge, CB3 OHA, UK}

\author[0000-0002-5082-5049]{Tong Bao}
\affiliation{INAF – Osservatorio Astronomico di Brera, Via E. Bianchi 46, 23807 Merate (LC), Italy}

\author[0000-0002-6211-7226]{Raimundo Lopes de Oliveira}
\affiliation{Departamento de F\'isica, Universidade Federal de Sergipe, Av. Marechal Rondon, S/N, 49100-000, S\~ao Crist\'ov\~ao, SE, Brazil}
\affiliation{Observat\'orio Nacional, Rua Gal. Jos\'e Cristino 77, 20921-400, Rio~de~Janeiro, RJ, Brazil}

\author[0000-0001-8722-9710]{Gavin Ramsay}
\affiliation{Armagh Observatory and Planetarium, College Hill, Armagh, BT61 9DG, N. Ireland, UK}

\author[0000-0003-4373-7777]{Paula Szkody}
\affiliation{Department of Astronomy, University of Washington, 3910 15th Avenue NE, Seattle, WA 98195, USA}

\author[0000-0002-3168-0139]{Matthew Graham}
\affiliation{Department of Astronomy, California Institute of Technology, 1200 East California Blvd, Pasadena, CA, 91125, USA}

\author{Thomas A. Prince}
\affiliation{Department of Astronomy, California Institute of Technology, 1200 East California Blvd, Pasadena, CA, 91125, USA}

\author[0000-0002-4770-5388]{Ilaria Caiazzo}
\affiliation{Institute of Science and Technology Austria (ISTA), Am Campus 1, 3400 Klosterneuburg, Austria}
\affiliation{Department of Astronomy, California Institute of Technology, 1200 East California Blvd, Pasadena, CA, 91125, USA}

\author[0000-0002-0853-3464]{Zachary P. Vanderbosch}
\affiliation{Department of Astronomy, California Institute of Technology, 1200 East California Blvd, Pasadena, CA, 91125, USA}

\author[0000-0002-2626-2872]{Jan van Roestel}
\affiliation{Anton Pannekoek Institute for Astronomy, University of Amsterdam, 1090 GE Amsterdam, The Netherlands}

\author[0000-0001-8372-997X]{Kaustav K. Das}
\affiliation{Department of Astronomy, California Institute of Technology, 1200 East California Blvd, Pasadena, CA, 91125, USA}

\author[0000-0003-3658-6026]{Yu-Jing Qin}
\affiliation{Department of Astronomy, California Institute of Technology, 1200 East California Blvd, Pasadena, CA, 91125, USA}

\author[0000-0002-5619-4938]{Mansi M. Kasliwal}
\affiliation{Department of Astronomy, California Institute of Technology, 1200 East California Blvd, Pasadena, CA, 91125, USA}

\author[0000-0002-9998-6732]{Avery Wold}
\affiliation{IPAC, California Institute of Technology, 1200 E. California
             Blvd, Pasadena, CA 91125, USA}

\author[0000-0001-5668-3507]{Steven L. Groom}
\affiliation{IPAC, California Institute of Technology, 1200 E. California
             Blvd, Pasadena, CA 91125, USA}

\author{Daniel Reiley}
\affiliation{Caltech Optical Observatories, California Institute of Technology, Pasadena, CA  91125}

\author{Reed Riddle}
\affiliation{Caltech Optical Observatories, California Institute of Technology, Pasadena, CA  91125}

\begin{abstract}
We present volume-limited samples of cataclysmic variables (CVs) and AM CVn binaries jointly selected from SRG/eROSITA eRASS1 and \textit{Gaia} DR3 using an X-ray + optical color-color diagram (the ``X-ray Main Sequence"). This tool identifies all CV subtypes, including magnetic and low-accretion rate systems, in contrast to most previous surveys. We find 23 CVs, 3 of which are AM CVns, out to 150 pc in the Western Galactic Hemisphere. Our 150 pc sample is spectroscopically verified and complete down to $L_X = 1.3\times 10^{29} \;\textrm{erg s}^{-1}$ in the 0.2--2.3 keV band, and we also present CV candidates out to 300 pc and 1000 pc. We discovered two previously unknown systems in our 150 pc sample: the third nearest AM CVn and a magnetic period bouncer. We find the mean $L_X$ of CVs to be $\langle L_X \rangle \approx 4.6\times 10^{30} \;\textrm{erg s}^{-1}$, in contrast to previous surveys which yielded $\langle L_X \rangle \sim 10^{31}-10^{32} \;\textrm{erg s}^{-1}$.  We construct X-ray luminosity functions that, for the first time, flatten out at $L_X\sim 10^{30} \;
\textrm{erg s}^{-1}$. We find average number, mass, and luminosity densities of $\rho_\textrm{N, CV} = (3.7 \pm 0.7) \times 10^{-6} \textrm{pc}^{-3}$, $\rho_M = (5.0 \pm 1.0) \times 10^{-5} M_\odot^{-1}$, and $\rho_{L_X} = (2.3 \pm 0.4) \times 10^{26} \textrm{erg s}^{-1}M_\odot^{-1}$, respectively, in the solar neighborhood. Our uniform selection method also allows us to place meaningful estimates on the space density of AM CVns, $\rho_\textrm{N, AM CVn} = (5.5 \pm 3.7) \times 10^{-7} \textrm{pc}^{-3}$. Magnetic CVs and period bouncers make up $35\%$ and $25\%$ of our sample, respectively. This work, through a novel discovery technique, shows that the observed number densities of CVs and AM CVns, as well as the fraction of period bouncers, are still in tension with population synthesis estimates.
\end{abstract}

\section{Introduction}
Cataclysmic Variables (CVs) are the most numerous accreting compact object binaries in the Milky Way. They serve as important laboratories of accretion physics and binary evolution which extend to many other types of binaries \citep[e.g.][]{2023tauris}. CVs are important in a broader astrophysical context since they likely dominate the hard X-ray excess seen from the Galactic Center \citep[GCXE;][]{2016hailey} and Galactic Ridge \citep[GRXE;][]{2006revnivtsev}. CVs are composed of a white dwarf (WD) accreting from a Roche-lobe filling donor, which is typically a late-type star \citep[e.g.][]{1995warner, hellierbook}. In magnetic CVs, where the WD magnetic field is strong enough that the Alfvén radius extends well past the surface of the WD, the disk is either substantially truncated (intermediate polars; $B\approx$ 1--10 MG) or entirely prevented from forming (polars; $B\approx$ 10--250 MG). In both cases, matter is channeled via magnetic field lines onto the WD surface, rather than flowing through the disk boundary layer onto the WD as in non-magnetic CVs \citep[e.g.][]{2017mukai}. 

AM Canum Venaticorum (AM CVn) binaries are the ultracompact analogs of classical CVs, which are highly evolved, having undergone one or two common envelope events \citep[e.g.][]{1995warner, 2010solheim, 2018ramsay}. As a result they have have helium-dominated donors and orbital periods in the range of 5--65 min \citep{2018ramsay}. Because of their short orbital periods, some AM CVns will be among the strongest sources of millihertz gravitational waves as seen by the upcoming Laser Space Interferometer Antenna \citep[e.g.][]{2004nelemans, 2024kupfer}. 

The most fundamental questions surrounding CVs and AM CVns can only be answered through population studies: 1) What is the true space density and total number of CVs and AM CVns in the Milky Way? \citep[e.g.][]{2015goliasch, 2018belloni, 2001nelemans, 2013carter}; 2) What fraction of CVs are magnetic, and what is the origin of their magnetism? \citep[e.g.][]{2020pala, 2021schreiber}; 3) Why is the mean mass of WDs in CVs $\approx 30\%$ higher than that of single WDs? \citep[e.g.][]{2020zorotovic, 2022pala}; 4) Where are all of the ``period bouncers" --- systems that have a degenerate, brown dwarf donor, and are predicted to constitute 40--70\% of all CVs? \citep[e.g.][]{2015goliasch, 2018belloni, 2020pala}; and 5) What role does magnetic braking play in angular momentum loss (AML), which drives the evolution of CVs and AM CVn progenitors? \citep[e.g.][]{1983rappaport, 2003andronov, 2022elbadry}. Regarding AM CVns specifically, their formation channel is a major question that has garnered recent interest \citep[e.g.][]{2021vanroestel, 2023sarkar, 2023belloni}.

The biggest problems with CV and AM CVn surveys are incompleteness and inhomogeneity: no single survey method, aside from large spectroscopic surveys, has been sensitive to all subtypes. Combining systems from various photometric and spectroscopic surveys as was done in the seminal work by \cite{2020pala} is possible, but requires hard work and leads to a selection function that is difficult to characterize \citep[e.g.][]{2020pala, 2021inight}. This has strongly encumbered estimates of completeness, which have led to contrasting estimates of space densities and the relative number of CVs in different evolutionary stages. A brief summary of CV evolution reveals which surveys are biased towards the discovery of different CV subtypes. 

CVs are ``born" when the companion star in a post common envelope binary (PCEB) fills its Roche lobe and starts mass transfer to the WD. Depending mainly on the nature of the donor star, most CVs are born at orbital periods of $P_\textrm{orb} = $ 6--10 hr \citep[e.g.][]{2011knigge}. From this point, down to $P_\textrm{orb}\approx$ 3--4 hr, mass transfer is high, reaching levels of $\dot{M} \approx 10^{-8} M_\odot \textrm{ yr}^{-1}$. Non-magnetic CVs in this evolutionary state are usually discovered as novalikes \citep[e.g.][]{2023inight}, while magnetic systems in this evolutionary state are typically X-ray bright ($L_X\sim 10^{31} - 10^{33} \textrm{ erg s}^{-1}$) intermediate polars (IPs) \citep[e.g.][]{2019suleimanov}. As mass transfer continues, AML is high, driven mostly by magnetic braking of the donor star, and the accretion disk remains permanently ionized. As CVs evolve to shorter periods, $\dot{M}$ is reduced, which leads to the onset of a thermal instability \citep{1976pringle, 2001lasota}. This is believed to be the cause of dwarf nova outbursts, which observationally manifest themselves as a 2--8 mag transient brightening at optical wavelengths lasting a few tens of days \citep[e.g.][]{hellierbook, 2023inight}. 

At $P_\textrm{orb}\approx$ 2--3 hr, there has been observational evidence and theoretical explanations for the (contested) existence of a ``period gap". It is believed that mass transfer halts, presumably due to the donor star becoming fully convective causing magnetic braking to turn off \citep[][]{1983gap, 2001gap, 2016gap, 2024gap}. AML due to gravitational wave radiation brings CVs back into contact at $P_\textrm{orb} \approx 2$ hr, and mass transfer rates remain high enough for dwarf nova outbursts to be seen until CVs approach the canonical orbital period minimum of $\approx 78$ min \citep{1983minimum, 2011knigge}. At this stage in their evolution, mass transfer rates reduce to $\dot{M} \approx 10^{-11}-10^{-10} M_\odot \textrm{ yr}^{-1}$, and the timescale required for the onset of the thermal instability at such low accretion rates is typically much longer than observational timescales. At such low mass transfer rates, the X-ray luminosity is also dramatically reduced, reaching values as low as $L_X\sim 10^{29} \textrm{ erg s}^{-1}$ \citep{2013reis}. Once enough mass is depleted from the donor, nuclear burning halts, and the donor becomes degenerate. This means that the donor radius \textit{increases} as mass is lost. This causes CVs to ``bounce" and evolve to longer periods at the canonical period minimum of $P_\textrm{orb}\approx78$ min, leading to these low mass transfer rate ($\dot{M} \approx 10^{-11} M_\odot \textrm{ yr}^{-1}$) systems with degenerate donors being called ``period bouncers". 

At this point, it is clear that CVs at different evolutionary stages have very different observed phenomenology, and have historically been detected in different ways: 1) X-ray surveys (especially hard X-ray surveys sensitive to $E \gtrsim 10$ keV) have mainly discovered magnetic CVs: IPs and nearby polars, though nearby non-magnetic CVs with high mass transfer rates have also been discovered this way \citep[e.g. using the ROSAT all-sky soft X-ray mission and the \textit{Swift}/BAT and \textit{INTEGRAL} hard X-ray missions;][]{1997verbunt, 2000schwope, 2022suleimanov}; 2) Optical photometric surveys have mainly discovered (non-magnetic) dwarf novae \citep[e.g. using the Catalina Real-time Transient Facility and Zwicky Transient Facility;][]{2014breedt, 2020szkody}; and 3) large spectroscopic surveys have mainly discovered low-accretion rate CVs near the orbital period minimum \citep[e.g. using the Sloan Digital Sky Survey;][]{2009gaensicke, 2011szkody, 2023inight}. This has led to present CV catalogs such as the Ritter and Kolb catalog \citep{2003ritter} and the International Variable Star Index (VSX) catalog\footnote{\url{https://www.aavso.org/vsx/index.php?view=search.top}} being comprised of systems from many different surveys, each with their own biases.

A similar story applies to AM CVns, though they evolve from short to long orbital periods given the degenerate nature of their donor stars. Short orbital period ($P_\textrm{orb} = 5-20$ min) systems do not undergo optical outbursts due to their high mass transfer rates, but are X-ray bright\footnote{Aside from the shortest period AM CVns,  HM Cnc and V407 Vul, the rest of the population is X-ray faint, reaching only $L_X\sim10^{30}-10^{31}\textrm{ erg s}^{-1}$ as seen from X-ray follow-up observations \citep[e.g.][]{2005ramsay, 2006ramsay, 2023begari}.} and have been identified since the ROSAT era \citep[e.g.][]{2002israel, 2018ramsay}. At intermediate periods ($P_\textrm{orb} = 20-50$ min), accretion disks in AM CVns are subject to thermal instabilities, leading to dwarf nova outbursts which enable easy identification through optical photometric surveys \citep[e.g.][]{2021vanroestel}. However, the majority of AM CVns are expected to be long-period ($P_\textrm{orb} = 50-65$ min) systems, since they remain at low mass transfer rates for $\sim$ few Gyr \citep[e.g.][]{2001nelemans, 2021wong}. These systems, like low accretion rate CVs, do not outbust frequently, and have only been identified through large spectroscopic surveys from SDSS \citep{2007roelofs, 2013carter} or through their eclipses in large optical photometric surveys \citep{2022vanroestel}.   

Here, we present the first volume-limited survey of CVs, including AM CVns, selected using a single tool which is sensitive to all subtypes. We construct volume-limited samples of CV candidates out to 1000 pc, 300 pc, and 150 pc. The 150 pc sample is confirmed with optical spectroscopy, complete down to $L_X = 1.3\times10^{29} \textrm{erg s}^{-1}$ in the 0.2--2.3 keV range, and the main focus of this paper. This survey is made possible using a tool recently presented by \cite{r24} --- an X-ray + optical color-color diagram dubbed the ``X-ray Main Sequence" which efficiently distinguishes accreting compact objects from coronally active stars. \cite{r24} applied this tool to discover new CVs in a crossmatch of the \textit{XMM-Newton} catalog and \textit{Gaia}, and recently, \cite{2024galiullin_chandra} incorporated this tool to discover new CVs in a crossmatch between the  \textit{Chandra} Source Catalog and \textit{Gaia}. Here, we construct the X-ray Main Sequence through a crossmatch of the recently released SRG/eROSITA eRASS1 soft X-ray survey of the Western Galactic Hemisphere \citep{2024merloni} with \textit{Gaia} Data Release 3 \citep[DR3;][]{2023gaia_dr3}.

In Section \ref{sec:xray}, we show that given a sensitive enough X-ray survey, CVs of all subtypes can be detected. In Section \ref{sec:sample}, we present the crossmatch between the eRASS1 catalog and \textit{Gaia} DR3 to identify optical counterparts. We outline the construction of CV samples out to 150, 300, and 1000 pc, as well as a reference sample of CVs created from the VSX catalog. In Section \ref{sec:results}, we compare the X-ray luminosity distributions of volume-limited samples to those of our VSX reference samples. We also present new space densities and X-ray luminosity functions based on the 150 pc sample. In that section, we also present the two new CVs discovered within 150 pc using our method and summarize global properties of all CVs in the 150 pc sample. Finally, in Section \ref{sec:discussion}, we discuss our sample completeness. We compare our inferred space densities of CVs and AM CVns to those of previous surveys as well as theoretical predictions.

\section{X-ray Emission from Cataclysmic Variables}
\label{sec:xray}

\textit{All} CVs are X-ray emitters. Assuming that half of the gravitational potential energy associated with accretion is radiated away as X-rays\footnote{This assumes the standard \cite{1973ss} picture that half of the energy is radiated away by the boundary layer, with the other half stored by the accretion disk, though the constant of 1/2 may increase up to 1 when the disk is truncated by a magnetic WD.}, the X-ray luminosity can be expressed as in \cite{1985patterson}:
\begin{gather}
\label{eq:xray}
    L_X = \eta \varepsilon_X\varepsilon_\textrm{out} \frac{1}{2}\frac{GM_\textrm{WD}\dot{M}}{R_\textrm{WD}} \nonumber\\ = 6.9\times 10^{30} \left(\frac{\eta}{0.1}\right)\left(\frac{\varepsilon_X}{0.3}\right)\left(\frac{\varepsilon_\textrm{out}}{0.5}\right)\left(\frac{M_\textrm{WD}}{0.8 M_\odot}\right)\times\\ \nonumber\left(\frac{\dot{M}}{10^{-10} M_\odot \textrm{ yr}^{-1}}\right)\left(\frac{R_\textrm{WD}}{0.0105 R_\odot}\right)^{-1} \textrm{ erg s}^{-1}
\end{gather}
where $L_X$ is the X-ray luminosity in the 0.2--2.3 keV range, $\eta$ is the radiative efficiency, $\varepsilon_X$ is the correction factor for observing X-rays only in the 0.2--2.3 energy band (a power-law model with $\Gamma=2$ is assumed to calculate the above value of 0.3), $\varepsilon_\textrm{out}$ is the fraction of X-rays that are emitted outward and not absorbed by the WD \citep[we adopt a value of 0.5 as in][]{1985patterson}, the additional factor of $1/2$ represents the fraction of gravitational potential energy associated with accretion radiated by the boundary layer, and $M_\textrm{WD}$ and $R_\textrm{WD}$ are the mass and radius of the WD \citep{2020bedard}, respectively, and $\dot{M}$ is the mass transfer rate. 

While there are uncertainties in this simple treatment, particularly in estimating $\eta$ and $\varepsilon_\textrm{out}$, it is clear that the lowest mass transfer rates reached by CVs, $\dot{M} \approx 10^{-11} M_\odot \textrm{ yr}^{-1}$, means that they can reach very low X-ray luminosities, on the order of $L_X\sim 10^{29} \textrm{ erg s}^{-1}$. This is supported by observational evidence, notably the work of \cite{2013reis}, where it was found that the mean X-ray luminosity of an inhomogenously selected sample of low accretion rate CVs discovered by the Sloan Digital Sky Survey \citep[SDSS;][]{2000sdss} reached that value\footnote{These values hold for CVs (i.e. WDs with a Roche-lobe filling donors), whereas wind-accreting systems such as low accretion rate polars (LARPs) have been shown to accrete at even lower rates, producing $L_X \lesssim8\times 10^{28} \textrm{erg s}^{-1}$ \citep[e.g.][]{2004szkody}.}.

The first all-sky X-ray survey that is sensitive to such low-$L_X$ systems is eRASS1, carried out by the eROSITA telescope aboard the \textit{SRG} spacecraft \citep{2021sunyaev, 2024merloni}. eRASS1 goes nearly six times deeper than the last all-sky soft X-ray survey, ROSAT \citep{1999rosat2}: eRASS1 and the 2RXS ROSAT catalog have flux limits of $F_X = 5\times 10^{-14} \textrm{erg s}^{-1}\textrm{cm}^{-2}$ and $F_X = 3\times 10^{-13} \textrm{erg s}^{-1}\textrm{cm}^{-2}$ in the 0.2--2.3 and 0.1--2.4 keV bands, respectively \citep{2024merloni, 2016boller}. Another advantage of the eROSITA telescope is the improved localization of sources compared to ROSAT, with a mean positional error of 4.7" and 15", respectively \citep{salvato2021, 2009agueros}. 

The added depth of SRG/eROSITA enables the construction of volume-limited, X-ray selected, complete CV samples for the first time. At the $L_X \sim 10^{29} \textrm{erg s}^{-1}$ luminosity limit, the 2RXS ROSAT catalog is only complete out to $\sim$50 pc, while the eRASS1 SRG/eROSITA catalog is complete out to $\sim$150 pc, given their respective flux limits. The ROSAT distance limit of 50 pc is not very useful, since the nearest CV to Earth, WZ Sge, at 48 pc, is the only known system within 50 pc, while 42 CVs have been found in the entire sky out to 150 pc \citep{2020pala}. Since all CV subtypes are X-ray emitters, and the X-ray Main Sequence selects all CV subtypes without any obvious bias \citep{r24}, this likely represents the most effective and complete way to create a CV catalog to date.

\section{Sample Creation}
\label{sec:sample}
\subsection{X-ray + Optical Crossmatching}
We aim to create a pure and complete sample of SRG/eROSITA X-ray sources associated with \textit{Gaia} (localized to milliarcseconds) optical counterparts. We first consider the distribution of the separation between X-ray and optical positions, $r$. Assuming that the distribution of both X-ray and optical positional measurements is Gaussian about the true position, the distribution of $r$ is the well-known Rayleigh distribution \citep[Equation \ref{eq:dist}; e.g.][]{salvato2021, 2022srgz, 2023czesla}:
\begin{gather}
    p(r; \sigma_\textrm{sep}) = \frac{r}{\sigma_\textrm{sep}^2} \exp\left(-\frac{1}{2}\frac{r^2}{\sigma_\textrm{sep}^2}\right)
\label{eq:dist}
\end{gather}
where $\sigma_\textrm{sep}$ is the Rayleigh parameter, equivalent to the value of the mode of the distribution (the mean is $\sqrt{\pi/2}$ times that value). Thus, we can use this distribution to characterize the probability that a given optical match is the true counterpart to an X-ray position. 

\subsection{Volume Limited Samples}
\label{sec:volume_lim}
We start with the eRASS1 ``Main" catalog released by the German (DE) consortium of SRG/eROSITA\footnote{\url{https://erosita.mpe.mpg.de/dr1/AllSkySurveyData\_dr1/Catalogues\_dr1/}}. The eRASS1 release by the DE consortium covers all detections by SRG/eROSITA covering the Western half ($180^\circ < \ell < 360^\circ$) of the Galactic sky in the first semester of operations (December 2019 -- June 2020). The entire catalog contains 903,203 unique X-ray sources down to an all-sky flux limit of approximately $F_X \sim 5\times 10^{-14} \;\textrm{erg s}^{-1}\textrm{cm}^{-2}$ in the 0.2--2.3 keV range as part of the ``Main" catalog. We find the mean positional uncertainty (\texttt{POS\_ERR}) of all sources in eRASS1 to be 4.5". 

We keep only point sources, and those with a $< 1\%$ chance of being a spurious detection by enforcing \texttt{EXT == 0} and \texttt{DET\_LIKE\_0 $>$ 10}, respectively. This reduces the sample size to 427,241 (46\% of the original catalog), and reduces the mean positional uncertainty to 3.5". We then crossmatch this catalog with all sources in \textit{Gaia} Data Release 3 \citep[DR3;][]{2023gaia_dr3} within 20", which gives us 1,850,697 unique optical sources. Not all of those optical sources are expected to be the true counterparts. We use such a large crossmatch radius in order to distinguish the Rayleigh distribution (of likely matches) from the distribution of chance alignments (which increases with $r$, and requires a large crossmach radius to be well-populated). We also remove X-ray faint objects, keeping only those with $\texttt{ML\_FLUX\_1} > 3 \times 10^{-14} \textrm{erg s}^{-1}\textrm{cm}^{-2}$ in the 0.2--2.3 keV range\footnote{The presence of these outliers is likely due to the increased eROSITA depth in the ecliptic poles \citep{2024merloni}. All such systems are beyond 150 pc and this small difference in sensitivity does not affect our final analyses with the 150 pc sample.}.  

In order to select Galactic sources with well-measured \textit{Gaia} parallaxes and proper motions, we enforce the following:
\begin{itemize}
    \item \texttt{pmra} /\texttt{pmra\_error > 5} 
    \item \texttt{pmdec} /\texttt{pmdec\_error > 5}
    \item \texttt{parallax\_over\_error > 3}.
\end{itemize}
We do not enforce a cut on the astrometric solution using the \texttt{RUWE} parameter, since this removes known CVs within 150 pc (see Section \ref{sec:150_full}). Rather, we adopt the classifier from \cite{2022rybizki} to remove spurious astrometric solutions. After some experimenting to recover known CVs within 150 pc (while not losing any known systems), we enforced the following cuts, which have some overlap with those used successfully in other surveys \citep[e.g.][]{2021el-badry}:
\begin{itemize}
    \item \texttt{fidelity\_v2 > 0.75}. This corresponds to a $< 25$ \% chance of a spurious astrometric solution for all sources.
    \item \texttt{dist\_nearest\_neighbor\_at\_least\_4\_brighter == 0}. This ensures that the nearest neighbor $\geq 4$ mag brighter is over 30 arcsec away\footnote{We note that this condition could be a bit too strict. Relaxing this condition to allow for $\geq 4$ mag neighbors up to 10 arcsec away introduces no new CVs into the 150 pc sample, but does introduce three false matches within a few arcsec of bright ($\gtrsim$ 8 mag stars). Applying this relaxed condition to the 300 pc sample does introduce four recently verified CVs (A. Pala, priv. comm.), albeit with six false matches near bright stars.}.
    \item \texttt{norm\_dG == NaN OR norm\_dG < -3}. This ensures that there are no nearby neighbors or no nearby bright stars which can contaminate the photometry of the source.  
\end{itemize}

\subsubsection{The X-ray Main Sequence}
\label{sec:xms}

\cite{r24} showed that an empirical cut in the plane of $F_X/F_\textrm{opt}$ (vertical axis) vs. \textit{Gaia} BP -- RP (horizontal axis) can efficiently select CVs (among other compact object binaries) from other stellar X-ray sources. The cut is defined by $\log_{10} (F_X/F_\textrm{opt})=(\textrm{BP -- RP})-3.5$. It was shown that virtually all of the CVs in the Ritter and Kolb catalog \citep{2003ritter}  with \textit{Gaia} and \textit{XMM-Newton} counterparts are above the cut, regardless of subtype and without requiring an extinction correction.

We compute the optical flux in the Gaia G band in the same way as in \cite{r24}, with $F_\textrm{opt} = 10^{0.4(m_\odot - m)}  L_\odot/ (4\pi (1 \textrm{ AU})^2)$, where $m_\odot = -26.7$ and $m$ is \texttt{phot\_g\_mean\_mag}. We take $F_X$ as the X-ray flux in the 0.2--2.3 keV eROSITA band, \texttt{ML\_FLUX\_1}.

In order to improve our selection of CVs, rather than just take all objects above the ``empirical cut" from \cite{r24}, we experimented with recovering a highly cleaned sample of CVs (well-vetted systems with known subtypes and orbital periods from photometry and/or spectroscopy) from the International Variable Star Index (VSX) catalog. We describe this procedure in Appendix \ref{sec:selection}. The outcome of this was the adoption of a ``modified cut":
\begin{itemize}
    \item $\log_{10} F_X/F_\textrm{opt} > \textrm{BP -- RP}-3.5$\\if $-0.3 < \textrm{BP -- RP} < 0.7$
    \item $\log_{10} F_X/F_\textrm{opt} > \textrm{BP -- RP}-3$\\if $0.7 < \textrm{BP -- RP} < 1.5$
\end{itemize}
We then enforce this ``modified cut" as described above, as well as the following:
\begin{itemize}
    \item BP--RP $<$1.5. This ensures that a large population of (falsely matched) M dwarfs does not enter the sample while excluding virtually no previously known systems. We elaborate further on this cut in Appendix \ref{sec:selection}.
    \item BP--RP $>$-0.3. This ensures that no presumably single hot WDs enter the sample. While these could be true matches, no known CVs are this blue (see Appendix \ref{sec:selection}).
\end{itemize}

Finally, in Appendix \ref{sec:selection}, we show that at all distances, after enforcing the cuts above, the population of CV candidates is well-modeled by a Rayleigh distribution with $\sigma_\textrm{sep} \approx 2.2"$. To ensure $>99\%$ completeness, we keep all systems within a separation of $r < 7.7" \;(3.5\sigma_\textrm{sep})$. We also keep all systems out to $r < 11" \;(5\sigma_\textrm{sep})$ that are isolated in the field (i.e. \textit{no} other optical sources are present within 30 arcsec). This set of cuts also ensures that all systems in the 150 pc spectroscopically verified sample by \cite{2020pala} are recovered in \textit{our} 150 pc sample (see next section). However, such a large separation does introduce some false positives (primarily with background active galactic nuclei), which we are able to remove after manual inspection in the 150 pc sample (see next section).  

\begin{figure*}
    \centering
    \includegraphics[width=\textwidth]{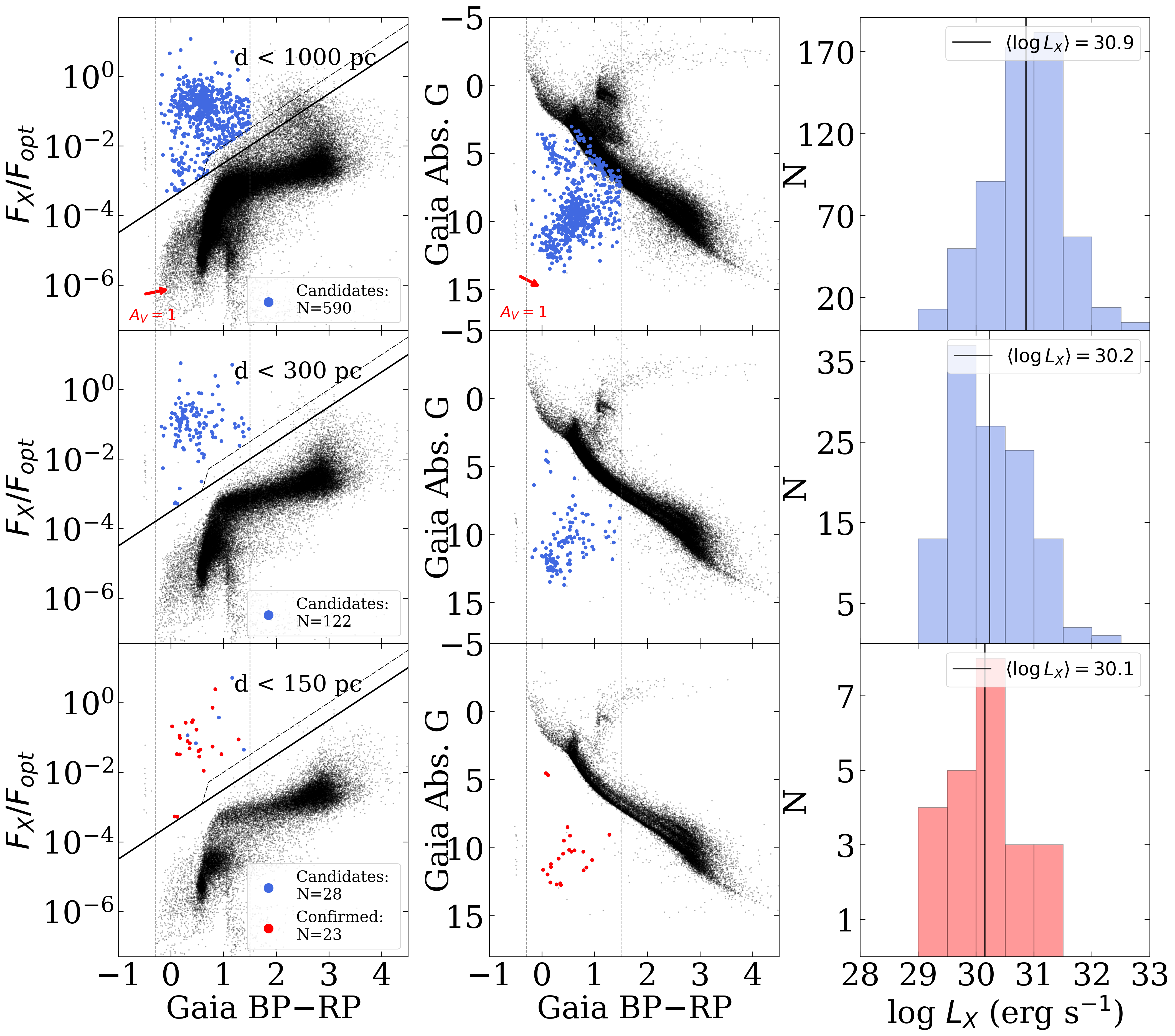}
    \caption{Volume-limited samples of CVs in eROSITA selected using the X-ray Main Sequence are shown out to distances of 150, 300, and 1000 pc (from bottom to top; tables available in machine-readable form). All objects in SRG/eROSITA eRASS1 within the given distance are shown in black, CV candidates in blue, and spectroscopically confirmed CVs in red. We select CVs using the dash-dot diagonal line in the X-ray Main Sequence (left), and exclude all objects with BP--RP $<-0.3$ (hot WDs) and BP--RP $>1.5$ (false matches). We plot CVs on an HR diagram constructed using all eRASS1 sources within each volume (center). The X-ray luminosity distribution out to each distance is shown on the right, with the 150 pc sample in red to show it is spectroscopically verified. }
    \label{fig:all_samples}
\end{figure*}

In Figure \ref{fig:all_samples}, we present our three volume-limited samples out to distances of 1000 pc, 300 pc, and 150 pc, created from the cuts described above. In the left panels, we show the X-ray Main Sequence constructed from all systems that meet the above cuts. Those that fall above the ``modified cut" and color cuts required to be a CV candidate are highlighted in blue. In the middle panels, those same systems are shown on a Hertzsprung-Russell diagram assembled from all systems that meet the cuts above, except for the ``modified cut" in the X-ray Main Sequence. The majority of CV candidates selected from the X-ray Main Sequence are located between the WD track and the main sequence, where they would be expected to reside. The right panels show luminosity distributions created from CV candidate systems, showing that the mean observed X-ray luminosity, $\langle \log L_X\rangle$, shifts depending on the survey volume. We describe each of the samples in more detail below. Elaborating on the point made by \cite{r24}, we also emphasize the remarkable efficiency of the X-ray Main Sequence in Table \ref{tab:volume_samples}. At all distance limits, $<1\%$ of sources are targeted as candidate CVs (or more generally, accreting compact objects) for spectroscopic follow-up.

\subsubsection{1000 pc and 300 pc Samples}
To create the 1000 pc sample, we enforced all cuts as described in Section \ref{sec:volume_lim} and \ref{sec:xms}, with the addition of \texttt{parallax > 1}. The total number of systems that pass all cuts is $N=590$, with an observed luminosity completeness of $L_X = 5.6 \times 10^{30} \textrm{erg s}^{-1}$ in the 0.2--2.3 keV range.

To create the 300 pc sample, we enforce \texttt{parallax > 1000/300}. The total number of systems that pass all cuts is $N=122$, with an observed luminosity completeness of $L_X = 5.1 \times 10^{29} \textrm{erg s}^{-1}$ in the 0.2--2.3 keV range. At distances beyond a few hundred pc, the Galactic hydrogen column, $N_H$, becomes non-negligible and is particularly efficient at absorbing soft X-rays. Therefore, at these distances, the true luminosity completeness starts to depend on the direction of the line of sight. However, because the luminosity limit of these two catalogues does not extend down to the lowest $L_X$ CVs, we do not explore these catalogs further in this work. 

However, we make these tables of CV candidates out to 300 pc and 1000 pc publicly available. We note that these samples likely have a high purity rate since they primarily contain objects between the WD track and main sequence on the \textit{Gaia} HR diagram (center panels Figure \ref{fig:all_samples}), though CVs with an evolved donor and/or bright disk may be near or on the main sequence \citep[e.g.][]{2020abril}. We are conducting a spectroscopic survey of the 1000 pc and 300 pc catalogs to investigate the demographics of more distant systems, which will be reported elsewhere. 

\subsubsection{150 pc Sample}
Finally, we created a 150 pc sample (\texttt{parallax > 1000/150}), which is the main focus of this paper\footnote{In order to best calculate \textit{Gaia} distances, we used the median posteriors from \cite{2021bailer}. Even after extending the parallax limit to  \texttt{parallax > 1000/175}, no new sources were found to be within 150 pc.}. We focus primarily on this sample for two reasons. Firstly, given the completeness of eRASS1 down to $F_X = 5\times 10^{-14} \textrm{erg s}^{-1}\textrm{cm}^{-2}$, the X-ray luminosity completeness of a 150 pc catalog is $L_X = 1.3 \times 10^{29} \textrm{erg s}^{-1}$. This is good agreement with the lower limit of $L_X$ observed in CVs in the past --- \cite{2013reis} intentionally targeted low accretion rate WZ Sge-like systems identified with SDSS, and only 1 of 20 systems (5 \%) in their sample has an X-ray luminosity below our completeness threshold (even after accounting for the flux difference in \textit{Swift}/XRT and SRG/eROSITA energy bands). In contrast, the luminosity completeness limit of a 300 pc sample constructed with eRASS1 would miss 6 out of 20 systems in the \cite{2013reis} sample (30\%).


\begin{table}
    \centering
    \begin{tabular}{c||c|c|c|c}
         \parbox{1.2cm}{\centering Volume Limited Sample Radius (pc)} & 
         \parbox{1.3cm}{\centering Complete to $L_X$ Limit (erg/s)} &  
         \parbox{1.45cm}{\centering Candidate Galactic X-ray sources} &  
         \parbox{1.5cm}{\centering Candidate CVs} &  
         \parbox{1.2cm}{\centering Confirmed CVs} \\
         \hline
         1000 & $5.6 \times 10^{30}$ & 73,129 & 590 (0.8\%) & {\parbox[c][1cm][c]{1.2cm}{\centering Future work}}\\
         \hline
         300 & $5.1 \times 10^{29}$ & 43,403 & 122 (0.3\%) & {\parbox[c][1cm][c]{1.2cm}{\centering Future work}}\\
         \hline
         150 & $1.3 \times 10^{29}$ & 23,718 & 28 (0.1\%) & 23
    \end{tabular}
    \caption{The X-ray Main Sequence efficiently selects CV candidates from other Galactic X-ray sources in volume-limited samples. Percentages in the fourth column are taken with respect to sources in the third. The 150 pc sample is the only one complete down to the low $L_X$ end of CVs.}
    \label{tab:volume_samples}
\end{table}


Our 150 pc catalog, constructed with the cuts described above, yields 28 candidate CVs. We visually inspect all systems, using the Aladin Lite \citep{aladin} tool to overlay sources from \textit{Gaia} DR3 on images from the PanSTARRS PS1, Digitized Sky Survey (DSS), or DESI Legacy DR10 surveys to search for possible nearby faint contaminants. We first eliminate two systems: 1eRASS J120125.8+084801 since it is $\sim 3"$ away from a known quasar and 1eRASS J170655.5-350631 since there are 6 other \textit{Gaia} sources within a $\sim 5"$ radius, a subset of which could dominate the observed X-ray flux at that position. Moreover, the \textit{Gaia} XP spectrum of that source shows Balmer absorption, but no emission lines. We also eliminate 1eRASS J035319.7-560057 since the DESI Legacy DR10 images reveal that the X-ray position is closer to an extended object $\sim 3"$ away, which is likely an active galaxy or cluster, and therefore the true counterpart to the X-ray source. Moreover, this \textit{Gaia} source is significantly fainter in the optical than the rest of the verified CVs, having a \textit{Gaia} G magnitude of $20.6$, whereas the rest of the verified CVs are more commonly between $G\approx 16 - 18$ mag.

After removing the above false matches, we are left with 25 candidate CVs. Of these, there are two true X-ray + optical matches, which are indeed compact object binaries as predicted by the X-ray Main Sequence, but not CVs. The first is 1eRASS J043715.9-471509, which is a known recycled pulsar + WD binary \citep[e.g.][]{1993danziger}. Previously, \textit{Chandra} observations revealed X-rays from this system \citep{2002zavlin}, yet it appears that the WD remains poorly studied, likely due to its optical faintness \citep{2012durant}. The second such system is AR Sco, the prototypical WD ``pulsar", in which a rapidly spinning magnetic WD generates synchrotron emission presumably through interactions with the magnetic field of its M dwarf companion \citep{2016marsh}. This system is not a CV, since no mass transfer is occurring, though it is close to filling its Roche lobe \citep{2016marsh, 2022pelisoli}. The recovery of these systems reveals the remarkable flexibility of the X-ray Main Sequence in revealing a breadth of compact object binaries.

After removing the above systems, we were left with 23 candidate CVs. Twenty-one of those systems are previously known CVs, and present in the \cite{2020pala} catalog. In Table \ref{tab:150_sample}, we use the classifications from \cite{2020pala}, with one exception --- SDSS J125044.42
+154957.3, which we describe further in Section \ref{sec:new_pb}. ``AM '' corresponds to (magnetic) polars, ``SU" corresponds to outbursting non-magnetic systems that show SU UMa outburst behavior \citep[i.e. superoutbursts;][]{1989su_uma}, ``UG" corresponds to outbursting non-magnetic systems that show U Gem outburst behavior (i.e. dwarf novae), ``NL" corresponds to non-outbursting non-magnetic systems in a high accretion state, ``UGWZ" corresponds to non-magnetic systems that only show outburst behavior on very long timescales (and appear quiescent otherwise) in a low accretion state like WZ Sge, ``IP" corresponds to intermediate polars, and ``AM CVn" corresponds to AM CVns. For further detail, we refer the reader to \cite{2020pala} for references to individual systems, and \cite{hellierbook} and \cite{2023inight} for modern reviews of subtypes and phenomenology. Of the 42 systems in the \cite{2020pala} catalog, 24 are located in the Western Galactic Hemisphere and should in principle be recovered by our analysis. However, V379 Vir is one of the three systems in \cite{2020pala} not recovered in our 150 pc sample, which an updated \textit{Gaia} DR3 parallax reveals to be located at a distance 155 $\pm$ 5 pc and beyond our 150 pc distance limit (\cite{2020pala} used \textit{Gaia} DR2). This means that we recover 21 of the 23 CVs within 150 pc in the Western Galactic Hemisphere reported by \cite{2020pala}, leaving Gaia J154008.28–392917.6 ($d=139$ pc) and V379 Tel  ($d=131$ pc) as the only systems in the \cite{2020pala} catalog not recovered by our analysis. We explore possible reasons for this and compare our 150 pc sample to that of \cite{2020pala} further in Section \ref{sec:prev_studies}.

This left 2 systems left to verify, both of which we confirm as CVs --- 1eRASS J054726.9+132649 is a new magnetic period bouncer and 1eRASS J101328.7-202848 is a new AM CVn. To help characterize those systems, we queried the optical time domain photometric database of the Zwicky Transient Facility \citep[ZTF;][]{bellm2019, graham2019, dekanyztf, masci_ztf}. ZTF is a 47 $\textrm{deg}^2$ field-of-view camera on the Samuel Oschin 48-inch telescope at Palomar Observatory, which scans the DEC $>$ --28$^\circ$ sky on a two-day cadence. Both new systems and the data associated with them are described in further detail in Section \ref{sec:new_systems}. In summary, our 150 pc catalog, after visual inspection and removal of other compact object binaries, is comprised of 23 \textit{spectroscopically confirmed} CVs, 2 of which are discoveries made possible thanks to the X-ray Main Sequence.

\begin{table*}
    \centering
\begin{tabular}{l||l|c|c|c|c|c|c|c}
\toprule
\parbox{3.9cm}{eROSITA Name} & \parbox{2.6cm}{Other Name} & \parbox{1cm}{$P_\textrm{orb}$ (min)} & \parbox{1.8cm}{$F_X$ $(\times 10^{13}\textrm{ erg}$ $\textrm{s}^{-1} \textrm{cm}^{-2})$} & \parbox{1.6cm}{Distance (pc)} & \parbox{1.5cm}{$L_X$ $(\times 10^{30}\textrm{ erg}$ $\textrm{s}^{-1})$} & \parbox{0.55cm}{Gaia G mag} & \parbox{0.55cm}{$r_\textrm{sep}$ (arcsec)} & \parbox{0.8cm}{Type} \\
\hline\hline
{\parbox[c][1cm][c]{3.9cm}{\centering 150 pc CVs + AM CVns}}  &&&&&&&&\\ \hline
1eRASS J014100.5-675326 & BL Hyi & 113.64 & $26.54\pm 1.15$ & $129.29_{-0.82}^{+0.83}$ & $5.02_{-0.37}^{+0.40}$ & 17.2 & 1.25 & AM \\
1eRASS J040911.6-711741 & VW Hyi & 106.85 & $35.17\pm 0.91$ & $53.74_{-0.05}^{+0.06}$ & $1.15_{-0.05}^{+0.05}$ & 13.8 & 1.42 & UGSU \\
1eRASS J054726.9+132649 & New (this work) & 93.948 & $0.91\pm 0.30$ & $132.06_{-2.66}^{+3.06}$ & $0.18_{-0.10}^{+0.10}$ & 18.3 & 3.38 & AM* \\
1eRASS J075505.1+220003 & U Gem & 254.74 & $68.94\pm 3.09$ & $93.06_{-0.34}^{+0.31}$ & $6.75_{-0.54}^{+0.50}$ & 13.9 & 1.26 & UG \\
1eRASS J080727.2-763200 & Z Cha & 107.28 & $8.08\pm 0.43$ & $120.23_{-1.13}^{+1.02}$ & $1.32_{-0.13}^{+0.11}$ & 15.7 & 1.80 & UGSU \\
1eRASS J081506.7-190318 & VV Pup & 100.44 & $5.22\pm 0.76$ & $135.26_{-0.72}^{+0.75}$ & $1.08_{-0.26}^{+0.26}$ & 16.0 & 0.97 & AM \\
1eRASS J081518.7-491324 & IX Vel & 279.25 & $28.95\pm 1.18$ & $90.31_{-0.14}^{+0.14}$ & $2.66_{-0.17}^{+0.19}$ & 9.3 & 4.26 & NL \\
1eRASS J100622.0-701404 & OY Car & 90.89 & $13.09\pm 0.55$ & $90.24_{-0.19}^{+0.24}$ & $1.21_{-0.08}^{+0.08}$ & 15.6 & 0.31 & UGSU \\
1eRASS J101328.7-202848 & New (this work) & - & $1.57\pm 0.44$ & $139.55_{-3.32}^{+2.86}$ & $0.34_{-0.16}^{+0.16}$ & 18.3 & 4.84 & AMCVn \\
1eRASS J105010.8-140438 & \parbox{2.6cm}{\vspace{5pt}\centering1RXS J105010.3 --140431} & 88.56 & $1.32\pm 0.41$ & $108.71_{-1.07}^{+0.97}$ & $0.17_{-0.08}^{+0.09}$ & 17.1 & 2.07 & UGWZ* \\
1eRASS J110539.3+250634 & ST LMi & 113.89 & $3.16\pm 0.64$ & $114.23_{-0.90}^{+0.90}$ & $0.47_{-0.15}^{+0.15}$ & 16.2 & 8.65 & AM \\
1eRASS J113826.8+032205 & QZ Vir & 84.7 & $32.65\pm 1.89$ & $125.65_{-0.74}^{+0.86}$ & $5.83_{-0.58}^{+0.56}$ & 15.9 & 1.27 & UGSU \\
1eRASS J115526.4-564151 & V1040 Cen & 87.11 & $119.06\pm 2.26$ & $127.43_{-0.18}^{+0.22}$ & $21.87_{-0.66}^{+0.70}$ & 14.0 & 2.11 & UGSU \\
1eRASS J125044.3+154952 & \parbox{2.6cm}{\vspace{5pt}\centering SDSS J125044.42 +154957.3} & 86.3 & $0.72\pm 0.27$ & $131.14_{-2.37}^{+2.27}$ & $0.14_{-0.10}^{+0.09}$ & 18.2 & 5.70 & AM* \\
1eRASS J125223.9-291454 & EX Hya & 98.26 & $444.97\pm 5.52$ & $56.77_{-0.05}^{+0.05}$ & $16.22_{-0.31}^{+0.32}$ & 13.2 & 3.83 & IP \\
1eRASS J130541.8+180103 & GP Com & 46.57 & $25.25\pm 1.41$ & $72.71_{-0.26}^{+0.25}$ & $1.51_{-0.15}^{+0.14}$ & 15.9 & 7.57 & AMCVn \\
1eRASS J131246.1-232133 & V396 Hya & 65.1 & $6.83\pm 0.69$ & $96.35_{-1.30}^{+1.57}$ & $0.71_{-0.12}^{+0.12}$ & 17.6 & 4.06 & AMCVn \\
1eRASS J140907.2-451716 & V834 Cen & 101.52 & $158.06\pm 3.35$ & $107.58_{-0.82}^{+0.89}$ & $20.66_{-0.85}^{+0.84}$ & 16.6 & 1.23 & AM \\
1eRASS J151955.1-250027 & GW Lib & 76.78 & $2.41\pm 0.39$ & $112.76_{-0.73}^{+0.72}$ & $0.34_{-0.08}^{+0.09}$ & 16.5 & 3.20 & UGWZ* \\
1eRASS J161514.8-283732 & V893 Sco & 109.38 & $12.74\pm 1.00$ & $116.52_{-0.49}^{+0.44}$ & $1.97_{-0.27}^{+0.24}$ & 15.0 & 1.43 & UGSU \\
1eRASS J170818.9-254833 & V2051 Oph & 89.9 & $2.22\pm 0.47$ & $107.98_{-0.50}^{+0.40}$ & $0.29_{-0.10}^{+0.10}$ & 15.4 & 2.91 & UGSU \\
1eRASS J194740.7-420026 & V3885 Sgr & 298.31 & $12.08\pm 1.25$ & $128.57_{-0.53}^{+0.56}$ & $2.26_{-0.34}^{+0.39}$ & 10.2 & 3.00 & NL \\
1eRASS J235301.0-385147 & BW Scl & 78.23 & $8.53\pm 0.96$ & $93.32_{-0.54}^{+0.55}$ & $0.84_{-0.15}^{+0.16}$ & 16.3 & 2.28 & UGWZ* \\
\hline\hline
{\parbox[c][1cm][c]{3.9cm}{\centering Other Compact Object Binaries}} &&&&&&&&\\ \hline
1eRASS J043715.9-471509 & PSR J0437-47 & 8208 & $10.57\pm 0.46$ & $142.08_{-10.84}^{+10.83}$ & $2.40_{-0.60}^{+0.69}$ & 20.3 & 1.64 & \parbox{1cm}{\vspace{5pt}\centering Pulsar + WD}  \\
1eRASS J162147.2-225311 & AR Sco & 213.6 & $12.04\pm 0.97$ & $122.77_{-0.38}^{+0.47}$ & $2.05_{-0.26}^{+0.27}$ & 14.6 & 1.90 & \parbox[c][1cm][c]{1cm}{\centering WD Pulsar} \\
\hline
\end{tabular}
\caption{Our 150 pc volume-limited sample, created through an X-ray + optical crossmatch of SRG/eROSITA eRASS1 and \textit{Gaia} DR3 (table with additional columns available in machine-readable form). All 25 systems that pass our cuts and visual inspection are shown, along with 23 confirmed CVs, 3 of which are AM CVns. We highlight newly confirmed systems and provide subtypes and orbital periods $(P_\textrm{orb})$ for previously known systems following \cite{2020pala}, where a volume-limited sample was created primarily from optically-identified systems. The star (*) symbol indicates a system is a period bouncer.}
    \label{tab:150_sample}
\end{table*}

\subsection{Reference Samples}
\label{sec:ref_catalog}
In order to highlight the importance of volume-limited samples as well as the power of the X-ray Main Sequence in generating an unbiased sample of CVs, we construct two reference samples to which to compare our volume limited samples: the International Variable Star Index (VSX) catalog of CVs crossmatched with ROSAT as well as the VSX catalog crossmatched with eRASS1. The VSX catalog contains well-vetted CVs confirmed either through a publication or a human expert verifying an amateur discovery.  However, CVs may be submitted from time-domain optical surveys, spectroscopic surveys, X-ray surveys, near-infrared surveys, and from both sample papers and individual object papers\footnote{\url{https://www.aavso.org/vsx/index.php?view=about.faq}}. The VSX catalog is representative of how CV catalogs have been created in the past, since it suffers from a selection function that is virtually impossible to characterize. Additionally, we emphasize that this catalog is likely not 100\% pure and can contain non-CV systems. We simply use this catalog as a \textit{reference} to which to compare our volume-limited catalogs, but encourage further exploration to vet the purity of this catalog. However, previous CV studies have successfully made use of systems from this catalog \citep[e.g.][]{2015ak, 2022abrahams}. 

\begin{figure*}
    \centering
    \includegraphics[width=0.9\textwidth]{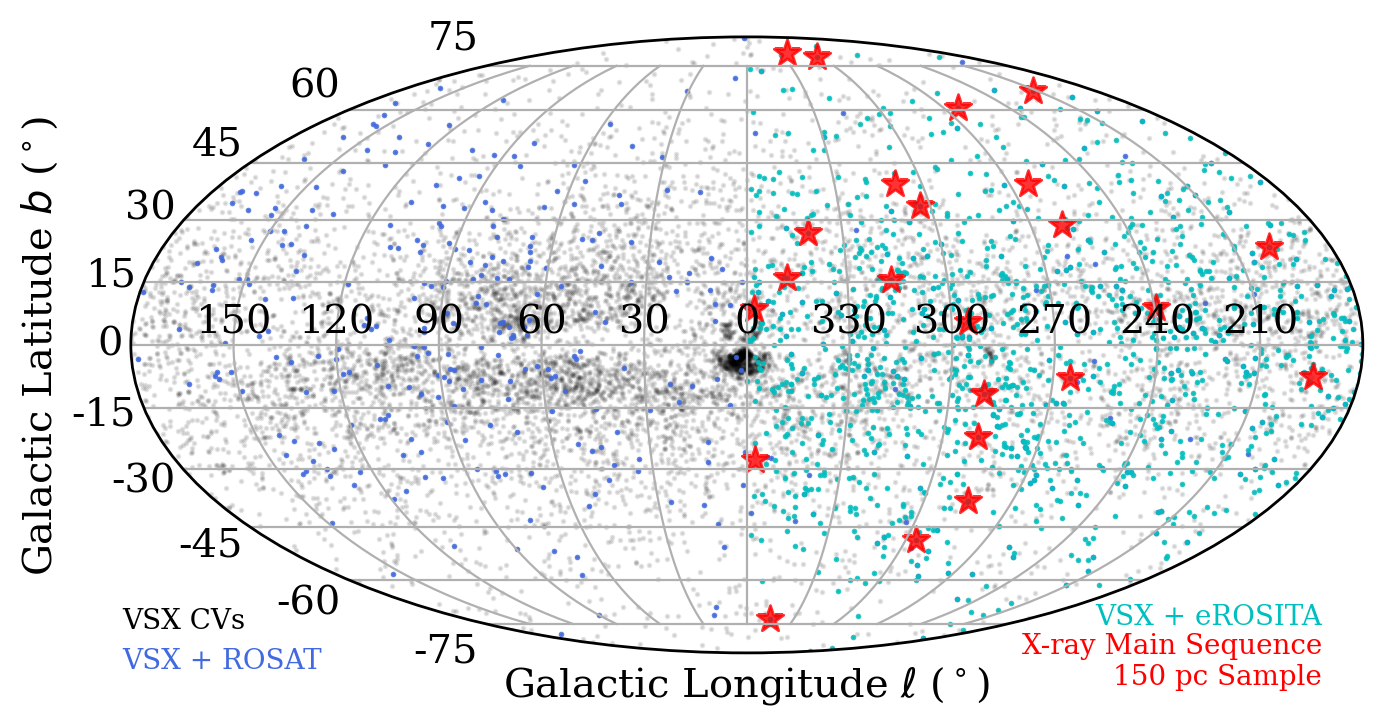}
    \caption{All-sky map of optically-selected CVs from the VSX database (black), VSX CVs with ROSAT detections (blue), and VSX CVs with eROSITA detections (cyan). We show our 150 pc sample of CVs identified by applying the X-ray Main Sequence to a crossmatch of SRG/eROSITA eRASS1 with \textit{Gaia} DR3 in red. The increase in density of X-ray detections from ROSAT (launched 1990) to SRG/eROSITA (launched 2019) is obvious. Catalogs assembled from optically-selected CVs suffer from a selection function that is difficult to characterize (e.g. the overdensity of CVs in the Galactic Bulge identified by OGLE).}
    \label{fig:map}
\end{figure*}

To create this catalog, we query the VSX database \footnote{\url{https://www.aavso.org/vsx/index.php}} for all CVs with the following classifications: AM (polar), DQ (intermediate polar), NL (novalike), ZAND (``Z And" types), UG (all ``U Gem" types including UGSU, UGER, UGSS, UGWZ, etc.), and CV (no subtype known). This yields a total of 14,756 systems. In order to eliminate some likely false positives, we eliminate all of those with an uncertain subtype, as indicated by the ``:" symbol in VSX. We are left with a total of 11,047 systems, which we show in Figure \ref{fig:map}. 

\begin{figure}
    \centering
    \includegraphics[width=0.45\textwidth]{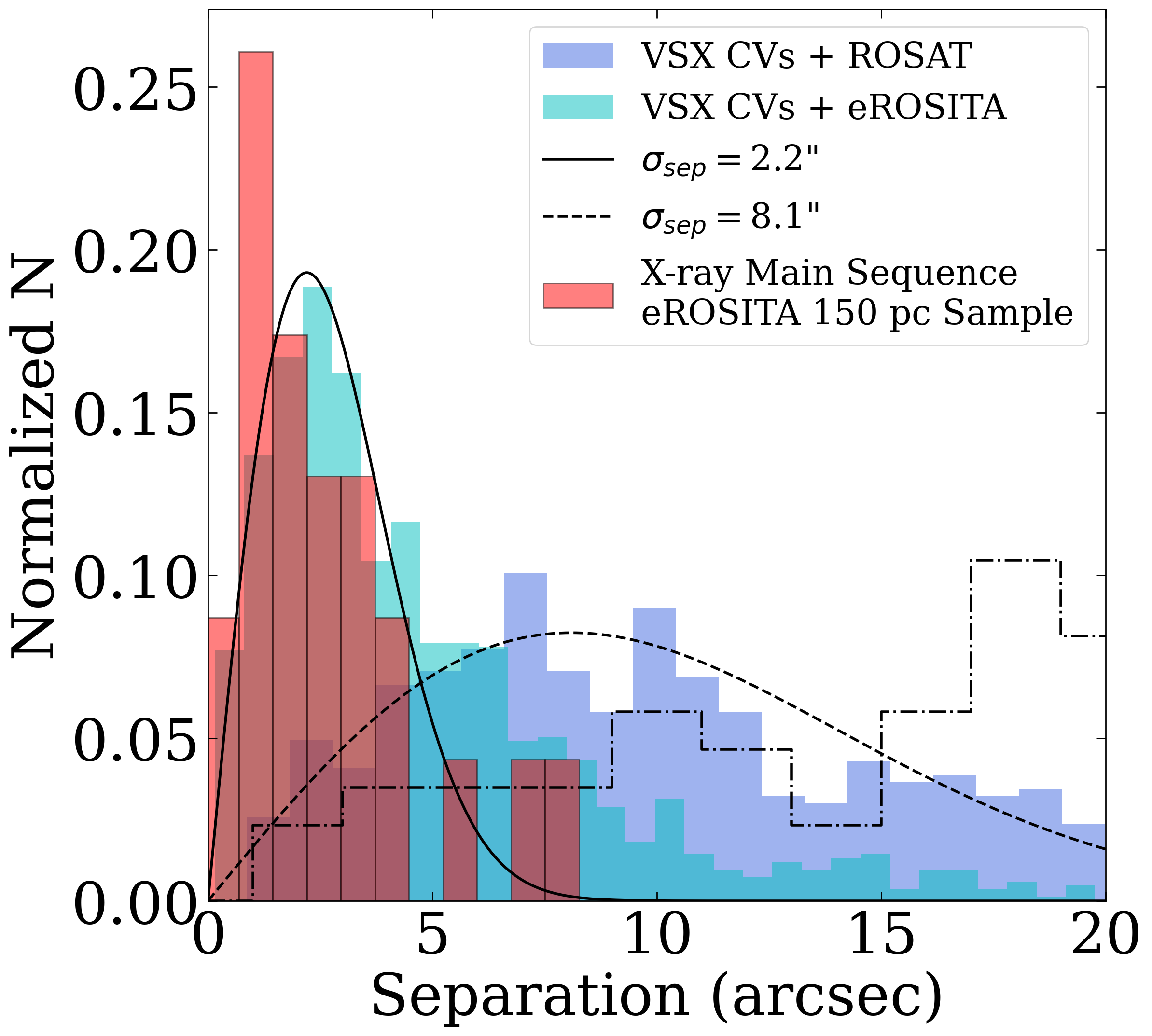}
    \caption{Distributions of $r$, the separation between the optical positions of VSX CVs and the nearest X-ray counterpart in the ROSAT 2RXS (blue) and SRG/eROSITA eRASS1 (cyan) catalogs. Our 150 pc sample is shown in red. Best-fit theoretical distributions (Equation \ref{eq:dist}) are shown as black dashed and solid lines, respectively. The distribution of $r$ for VSX optical points shifted by 2' is shown as a dash-dot line, demonstrating that the VSX CV + eROSITA catalog is dominated by true associations.}
    \label{fig:dist_main}
\end{figure}

We show the distribution of the separation between the X-ray and optical positions, $r$, of all sources in the VSX catalog crossmatched against the ROSAT 2RXS and SRG/eROSITA eRASS1 catalogs in Figure \ref{fig:dist_main}. We also shift all VSX positions by 2 arcmin, repeat the crossmatch with eRASS1, and plot the distribution of $r$ with the dash-dot line (false matches). To determine if an eROSITA point is a true crossmatch with a VSX CV, we require that $r < 3\sigma_\textrm{sep, eROSITA} = 6.6"$. According to Equation \ref{eq:dist}, this criterion ensures that $\sim 98\%$ of real associations are selected, assuming $r$ indeed follows a Rayleigh distribution. Beyond this separation, Figure \ref{fig:dist_main} shows that over half of potential crossmatches are more likely to be chance alignments. To construct our reference sample of VSX + ROSAT crossmatches, because of the larger positional error of ROSAT, we adopt a more conservative separation of  $r < 1.5\sigma_\textrm{sep, ROSAT} = 12.15"$. According to Equation \ref{eq:dist}, this criterion ensures that $\sim 67\%$ of real associations are selected. 

This exercise alone demonstrates the advantage of SRG/eROSITA over ROSAT. Using the criteria above, we find 983 VSX CV systems with an SRG/eROSITA counterpart, whereas only 134 systems have one in ROSAT. In order to create meaningful X-ray luminosity distributions in the following section, we require systems to have a well-measured (3$\sigma$) \textit{Gaia} parallax. This leaves 668 VSX CVs with an SRG/eROSITA counterpart and 129 systems with a ROSAT counterpart. To have a fair comparison with the public release of SRG/eROSITA eRASS1, we only reported here the number of ROSAT matches in the Western Galactic Hemisphere.

\section{Population Properties of the 150 pc Sample}
\label{sec:results}

\subsection{General Trends by CV Subtype}
\label{sec:150_full}

\begin{figure*}
    \centering
    \includegraphics[width=0.35\textwidth]{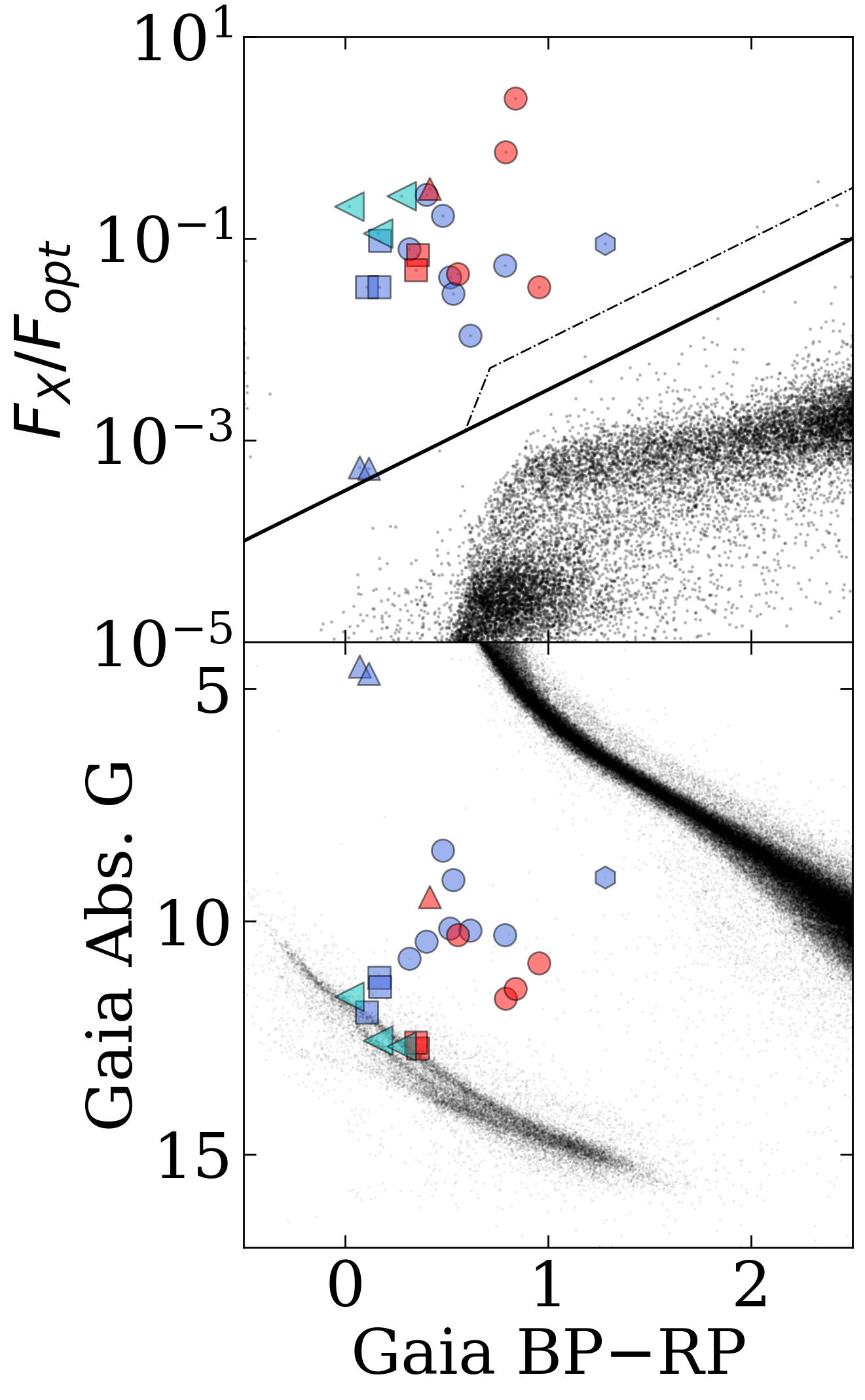}\includegraphics[width=0.65\textwidth]{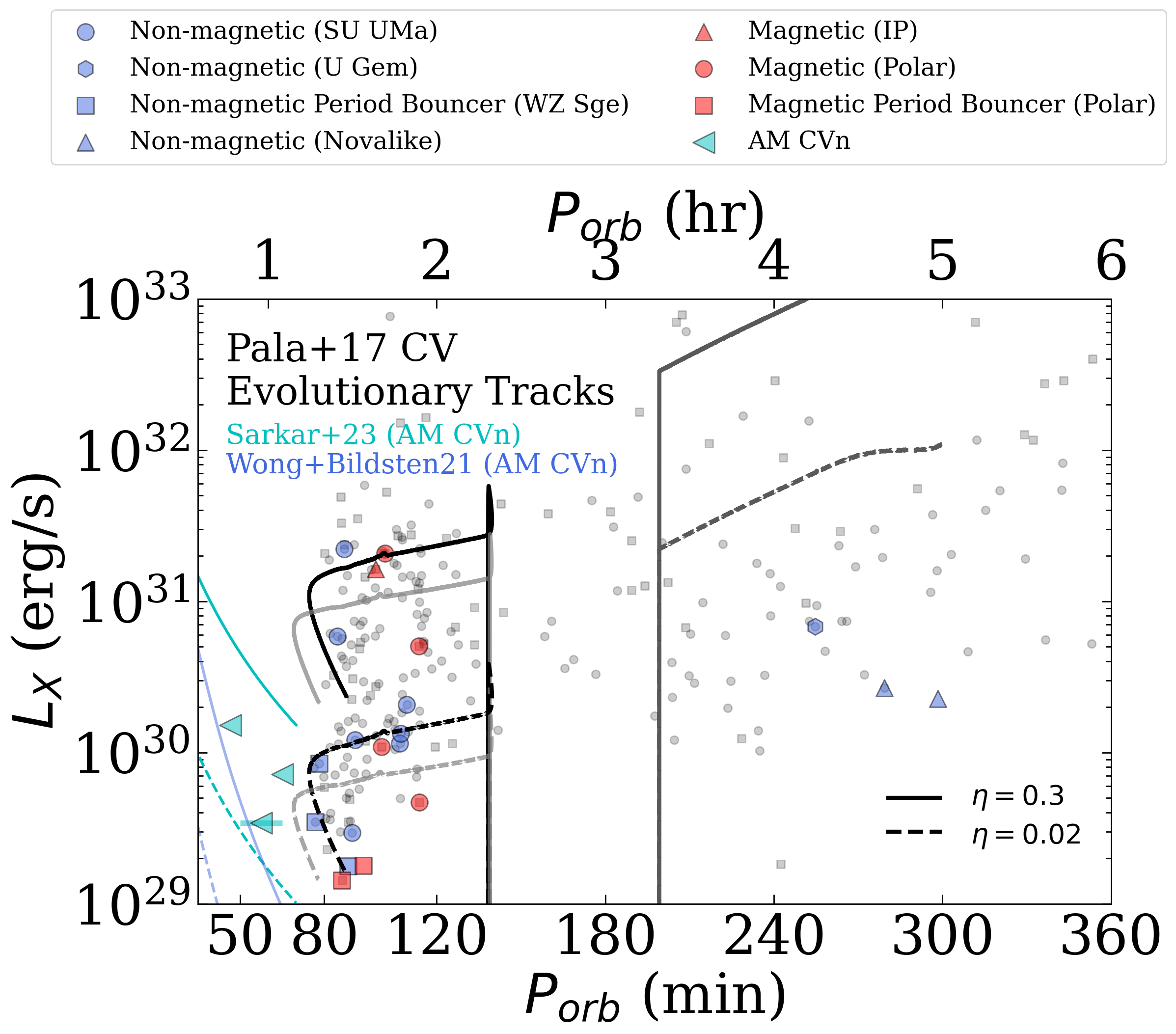}
    \caption{Our 150 pc sample in the X-ray Main Sequence (upper left, with all systems in our 150 pc crossmatch in black), \textit{Gaia} HR diagram (lower left, with the 100 pc \textit{Gaia} catalog in black), and in the $P_\textrm{orb}$--$L_X$ parameter space (right, with the VSX + eROSITA CV sample in gray). Theoretical $L_X$ values are plotted as a function of period, using donor mass loss estimates from the CV evolutionary tracks (``standard" in gray and ``optimal" in black) of \cite{2017pala},  ``He WD progenitor"/``evolved CV progenitor" AM CVn tracks of \cite{2021wong}/\cite{2023sarkar} and Equation 1. Accretion efficiencies in the range of $\eta \approx 0.02-0.3$ (dashed and solid tracks) best fit the data, which could be confirmed through X-ray spectral analyses that likely depend on CV subclass. Observations of trends in these parameter spaces by CV subclass are presented in Table \ref{tab:all}.}
    \label{fig:all}
\end{figure*}

We first present the global properties of the 150 pc sample. In Figure \ref{fig:all}, we split all CVs into subclasses using different symbols, and into magnetic, non-magnetic, and AM CVns using different colors. We summarize all properties of CVs based on subclass in Table \ref{tab:all}. In Figure \ref{fig:all}, we plot the orbital period of CVs in our sample versus the X-ray luminosity. In gray, we show systems from the VSX + eROSITA sample that have well-measured periods. We show magnetic systems (IPs and polars) as squares and non-magnetic systems as circles.

To provide a comparison to theoretical $L_X$ estimates, we overlay the CV evolutionary tracks by \cite{2017pala}\footnote{\url{https://zenodo.org/records/2592806}} in Figure \ref{fig:all}. Those tracks produce the donor mass loss rate\footnote{Especially in long period systems (above the period gap), there is some uncertainty in these estimates that arises from uncertainties in the strength of magnetic braking \citep[e.g.][]{2022elbadry}.} as a function of orbital period, and are effectively an updated version of the classic \cite{2011knigge} tracks, but created using Modules for Experiments in Stellar Astrophysics \citep[MESA;][]{2011mesa, 2013mesa, 2015mesa, 2018mesa}. We call the track that assumes AML only due to gravitational radiation below the period gap ``standard" (gray) and that which includes additional AML due to magnetic braking below the period gap ``optimal", following the convention of \cite{2011knigge}. These tracks have been computed assuming $M_\textrm{WD} = 0.8 M_\odot$ (unlike the tracks of \cite{2011knigge}, which assume $M_\textrm{WD} = 0.7 M_\odot$), $R_\textrm{WD} = 0.0105 R_\odot$ \citep{2020bedard}, and convert to X-ray luminosity in the 0.2--2.3 keV range following Equation \ref{eq:xray}. We plot two versions of each track, one assuming an accretion efficiency of $\eta = 0.3$ (solid lines) and another assuming $\eta = 0.02$ (dashed lines), which roughly encompass the X-ray luminosity of all CVs within our sample. However, we note that a different choice of $\eta_X$ (which depends on the X-ray spectral profile of a CV) and $\eta_\textrm{out}$ (which depends on the geometry of a CV) could lead to a different range of accretion efficiencies. For now, we show that a range of accretion efficiencies $\eta \approx 0.02-0.3$ reproduces the observed X-ray luminosity of CVs in our sample, but defer more detailed studies to future work. 

We also provide a comparison to theoretical $L_X$ estimates of AM CVns in Figure \ref{fig:all}. We estimate $L_X$ under the same assumptions as in Equation \ref{eq:xray}, taking the donor mass loss rates from the evolutionary tracks of \cite{2021wong} and \cite{2023sarkar}, adopting $M_\textrm{WD} = 0.75M_\odot$ and $M_\textrm{WD} = 0.8M_\odot$ and the corresponding $R_\textrm{WD}$ from \cite{2020bedard}, respectively. \cite{2021wong} (blue lines in Figure \ref{fig:all}) carried out AM CVn evolutionary models based on the He WD channel, where two common envelope phases lead to the creation of two WDs that commence accretion and lead to an AM CVn binary. We present the highest entropy model from that work, assuming a donor WD initial central temperature of $T_c = 3\times 10^7$ K. \cite{2023sarkar} (cyan lines in Figure \ref{fig:all}) carried out AM CVn evolutionary models based on the evolved CV channel, where a single common envelope phase leads a WD accreting from an evolved donor star, draining its hydrogen envelope and leading to the creation of an AM CVn binary. We present the most hydrogen exhausted model from that work, assuming a donor WD initial central temperature of $T_c = 3\times 10^7$ K. As with the CV evolutionary tracks, we plot the expected $L_X$ assuming efficiencies of $\eta=0.3$ (solid lines) and $\eta = 0.02$ (dashed lines). We see that the ``He WD progenitor" tracks of \cite{2021wong} underestimate long period AM CVn $L_X$ values, and that the ``evolved CV progenitor" AM CVn tracks of \cite{2023sarkar} match the observed $L_X$ values under reasonable assumptions of accretion efficiency. This could provide additional observational basis for recent claims made that most long-period AM CVns are created through the evolved CV channel (though short period systems could more likely originate from the He WD or He star channels), though more detailed analysis of more systems is needed to make such a claim \citep[e.g.][]{2023sarkar, 2023belloni}.

In Table \ref{tab:all}, we present general trends seen in X-ray luminosity and orbital period by CV subtype. We also list where they tend to be located in the X-ray Main Sequence and HR diagram, if there are sufficient systems to make such a claim. Some of the obvious trends are the clustering of novalikes away from the rest of the CVs in both the X-ray Main Sequence and the HR diagram, owing to the bright disk in these systems. Also robust is the location of (long period) AM CVns and period bouncers up against the WD track in the HR diagram. These systems are also located furthest in the upper left corner of the X-ray Main Sequence, due to the minimal contribution from the disk and donor. Generally, long period AM CVns and period bouncers also have the lowest X-ray luminosities of the entire sample, with all systems having $L_X \lesssim 2\times 10^{30} \textrm{erg s}^{-1}$ and four of the five period bouncers having $L_X \lesssim 3\times 10^{29} \textrm{erg s}^{-1}$.

We also note the obvious lack of high-$L_X$ ($\sim10^{33} \textrm{ erg s}^{-1}$) IPs in our sample --- only EX Hya, the closest IP to Earth is present. There has been some attention paid to so-called ``low-luminosity IPs" (LLIPs), which have lower X-ray luminosities than those found abundantly in hard X-ray surveys 
\citep{2014pretorius, 2023mukai}. EX Hya, like other LLIPs, has an X-ray luminosity of $\sim 10^{31} \textrm{erg s}^{-1}$. Based on the catalog of LLIPs available on the website of K. Mukai\footnote{\url{https://asd.gsfc.nasa.gov/Koji.Mukai/iphome/catalog/llip.html}}, it is clear that LLIPs tend to be located below the period gap ($P_\textrm{orb}\lesssim2$ hr) and at much closer distances ($d\lesssim500$ pc) compared to their high-luminosity counterparts ($P_\textrm{orb}\gtrsim4$ hr;  $d\gtrsim1$ kpc). This could signal that LLIPs are actually the dominant population of IPs and that the 300 pc and 1000 pc samples we present here could harbor several previously unknown LLIPs.

There is also a lack of short-period ($\lesssim 50$ min) AM CVns. These systems are believed to be rare, with AM CVns that have $P_\textrm{orb}\lesssim 20$ min existing for only $\sim$0.1 Gyr, those that have $P_\textrm{orb}\lesssim 50$ min existing for only $\sim$1 Gyr and long period systems ($P_\textrm{orb}\gtrsim 50$ min) existing up to a Hubble time \citep[e.g.][]{2021wong}. Therefore, it is not surprising that we only find long-period systems in our sample. This means that our survey indeed targets the dominant population of AM CVns \citep[e.g.][]{2013carter, 2022vanroestel}. Most these systems have been difficult to find in the past and are virtually invisible, both optically and in X-rays, beyond a few hundred pc, unless discovered through rapid follow-up after a transient outburst \citep[e.g.][]{2021vanroestel}.

\begin{table*}
    \centering
    \begin{tabular}{l||l|l|l|l}
        \parbox[c][1.5cm][c]{5cm}{\centering CV Subtype \\(23 total systems)} &
        \parbox[c][1.5cm][c]{1.8cm}{\centering Typical \\$L_X$ (erg/s)} &
        \parbox[c][1.5cm][c]{1.8cm}{\centering Typical \\$P_\textrm{orb}$} &
        \parbox[c][1.5cm][c]{3cm}{\centering X-ray Main \\Sequence Location} &
        \parbox[c][1.5cm][c]{3cm}{\centering HR Diagram \\Location} \\
        \hline \hline
         Non-magnetic (16)&&&&\\ \hline
         AM CVn (3) & $10^{29} - 10^{30}$&50--60 min&Uppermost left&On WD track \\
         SU UMa-type (7) & $10^{29} - 10^{31}$&78 min--2 hr&Scattered&Between WD and MS tracks\\
        U Gem-type (1) & $\sim10^{31}$&$\gtrsim 4$ hr&Right&\parbox[c][2.5cm][c]{3cm}{ Between WD and MS tracks, closer to MS track}\\
         WZ Sge-type Period Bouncer (3) & $10^{29} - 10^{30}$&78--110 min hr&Middle Left&On or just above WD track\\
         Novalikes (2) &$\sim10^{30}$&4--5 hr&Lower left&High Abs. G, near MS track
         \\\hline
         Magnetic (7) &&&&\\ \hline
         Intermediate Polar (1) &$\gtrsim10^{31}$&-&-&Between WD and MS track\\
         Polar (4) &$10^{29}-10^{31}$&78 min--2 hr&Scattered/upper center&\parbox[c][2.5cm][c]{3cm}{ Between WD and MS track, redder than non-magnetic CVs}\\
         Polar Period Bouncer (2)&$\sim 10^{29}$&78--95 min&Center&On WD track
         \\ \hline
    \end{tabular}
    \caption{Summary of CV subtypes and observed properties. If there are not enough systems to make a claim, or if the single system is not representative of the overall population, we denote it with a ``-" symbol. Also presented are the location in the X-ray Main Sequence (within the upper left corner) and on the HR diagram. }
    \label{tab:all}
\end{table*}

\subsection{X-ray Luminosity Distributions}
\begin{figure}
    \centering
    \includegraphics[width=0.45\textwidth]{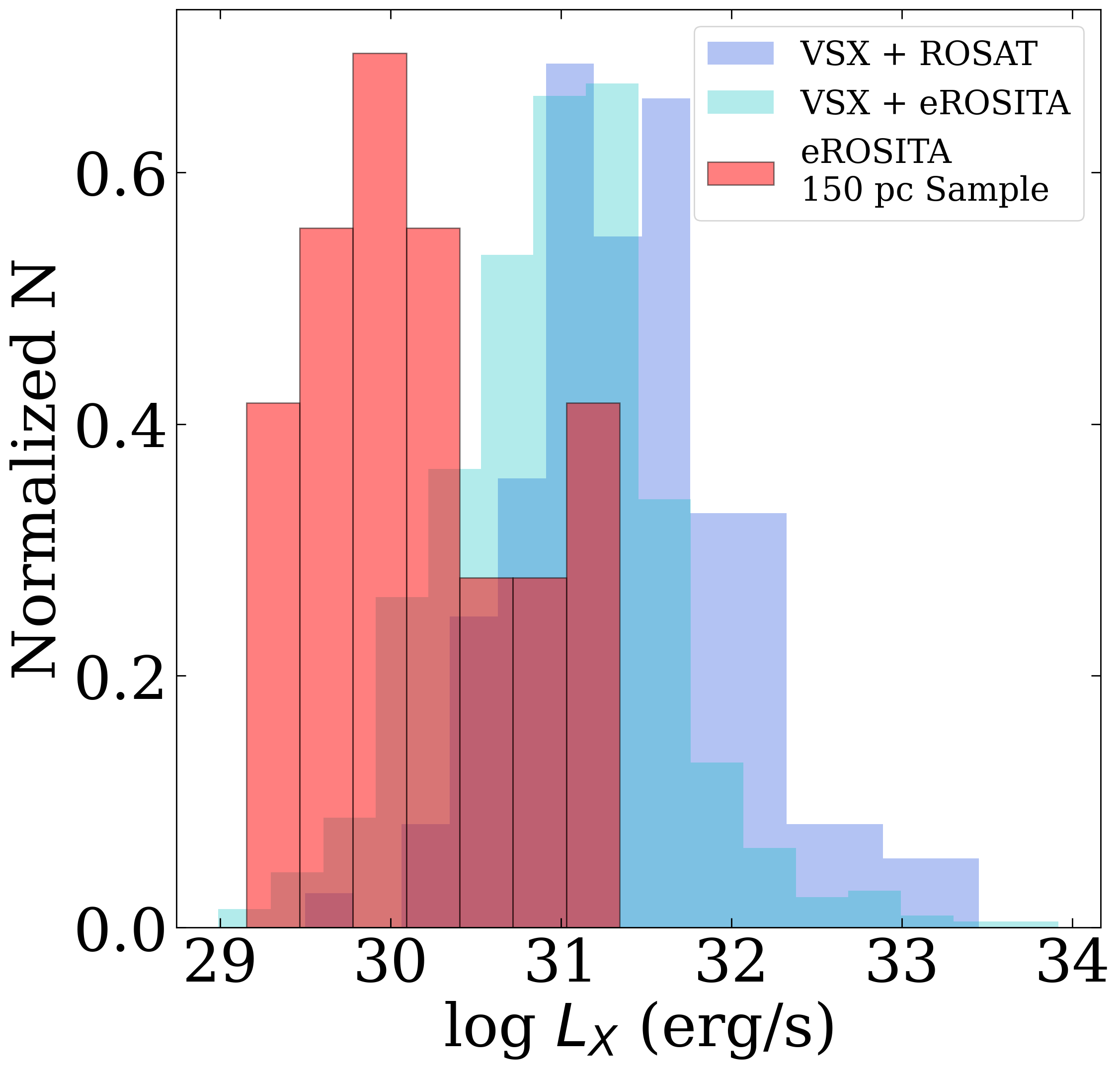}
    \caption{Observed $L_X$ distributions of CVs. $L_X$ distributions of primarily optically-identified systems that have eROSITA (cyan) and ROSAT (blue) X-ray counterparts suggest that CVs have $\langle L_X\rangle \sim 10^{31-32}\textrm{erg s}^{-1}$. Our 150 pc volume limited sample of systems selected using the X-ray Main Sequence (red) instead reveals that $\langle L_X \rangle \sim 10^{30}\textrm{erg s}^{-1}$.}
    \label{fig:lum_dist}
\end{figure}

We present the \textit{observed} X-ray luminosity distribution of three samples: the VSX + eROSITA, VSX + ROSAT, and our 150 pc sample, in Figure \ref{fig:lum_dist}. These are observed luminosity distributions because the X-ray fluxes are \textit{not} corrected for Galactic hydrogen column density, $N_H$. The 150 pc sample can safely be treated as the \textit{true} X-ray luminosity distribution, though, since the proximity of those systems means that the intervening Galactic hydrogen column density, $N_H$, is minimal. However, we note that detailed X-ray modeling of CVs has revealed intrinsic absorption, exceeding that predicted by optical/UV extinction estimates \citep[e.g.][]{2017mukai}. That would mean that \textit{every} CV would require X-ray spectral modeling to determine its true $L_X$, which is beyond the scope of this paper. Given this caveat, we estimate the maximum difference between the true and observed luminosities of the 150 pc sample to be a factor of $<0.07$. Based on the Bayestar19 map \citep{2019green}, even the most distant systems in the 150 pc sample have an upper limit of $A_V < 0.03$ mag, which corresponds to $N_H < 6 \times 10^{19} \textrm{cm}^{-2}$ \cite{2009guver}. We use the WebPIMMS tool \citep{1993pimms} to calculate the true flux of those systems, assuming a power law model with index $\Gamma \sim 2$. This leads to a difference between observed and true flux of $0.07$, though assuming harder X-ray spectra (either through power law models with lower values of $\Gamma$ or thermal bremsstrahlung models with $kT\sim 10$ keV) leads to even smaller differences ($<0.03$). 

We do not attempt to correct the observed luminosity distributions of the VSX samples, though, since this would require X-ray spectral fitting of each system. This would be difficult for the majority of systems that are in the low-count regime. We used the Bayestar19 \citep{2019green} map to approximate the value of $A_V$ for some of the most distant systems, along with the WebPIMMS tool to calculate the difference between the true and observed flux. We find that the most distant systems would require at most a correction of 2--3 in the 0.2--2.3 keV range. With most systems in the VSX sample located at a few hundreds of pc, we can safely assume that the observed luminosity distributions are no more than a factor of $\sim 2$ offset from the true distribution, though further work would be needed to clarify this. 

\subsection{X-ray Luminosity Functions and Space Densities}

\subsubsection{Galactic Model}
Three space densities can fully describe the global properties of CVs in the Milky Way, and can be used to construct luminosity functions  --- space densities as a function of X-ray luminosity. The number density, $\rho_N$, is the total number of systems in a given effective volume\footnote{An ``effective" volume assumes a density profile (not necessarily uniform as in a sphere), typically modeled as exponential for the disks of the Milky Way.} (units of $\textrm{pc}^{-3}$). The mass density, $\rho_M$, is the total number of systems per mass contained in a given effective volume (units of $M_\odot^{-1}$). The luminosity density, $\rho_L$, is the average X-ray luminosity per mass contained in a given volume (units of $\textrm{erg s}^{-1}M_\odot ^{-1}$).

We closely follow the method of \cite{2022suleimanov}, who calculated luminosity functions of hard X-ray emitting CVs. Though that work focused on flux-limited surveys, the method can be applied directly (and more easily) to volume-limited surveys. \cite{2022suleimanov} integrated over the entire Galactic mass distribution \citep{2002grimm}, whereas other works \citep[e.g.][]{2012pretorius, 2018schwope} use the $1/V_\textrm{max}$ method outlined by \cite{1993tinney}. This method was developed for flux-limited surveys on photographic plates, which assumed all objects were observed at approximately the same Galactic latitude\footnote{Though it is noted in \cite{2012pretorius} that the ROSAT RBS survey that was used observed in slices of $b$, and that they compute the corresponding $1/V_\textrm{max}$ for each slice.}. Because we enforce a volume limit, the integration over the full Galactic mass distribution is straightforward to do numerically. As in \cite{2022suleimanov}, we adopt a modern estimate of the Galactic stellar mass distribution computed by \cite{2016barros}. 

In order to compute the above space densities, we first calculate the amount of mass $\delta M$ contained in the volume of the Milky Way that we probe (i.e. in our case, how much mass is contained within 150 pc):
\begin{gather}
    \delta M = \int d\ell\int d \cos b \int \rho(z, R) \;r^2 dr
\end{gather}
where $\ell, b$ are Galactic coordinates\footnote{Note the change in coordinates needed for the spherical integral since Galactic $b$ is measured from the Galactic equator, unlike the typical $\theta$ used in spherical coordinates which is measured from the North pole.}, $z$ is the height above the Galactic plane, $R$ is the distance from a system to the Galactic Center, and $r$ is the distance from a system to the Sun. Since we only work with the Western Galactic Hemisphere, we integrate between $0 < \ell < \pi$ and $-1 < \cos b < 1$. We integrate from $0 < r < d_\textrm{max}$, where $d_\textrm{max}$ is the limiting distance of our survey (150 pc in our case). $\rho(z, R)$ is the mass density profile as calculated by \cite{2016barros}:
\begin{gather}
    \rho(z, R) = \frac{\Sigma_0}{2h}\exp\left[-\left(\frac{R-R_0}{R_d} + R_\textrm{ch}\left(\frac{1}{R} - \frac{1}{R_0}\right) + \frac{|z|}{h}\right)\right]
    \label{eq:mass}
\end{gather}
where $\rho (z, R)$ has units of $M_\odot \textrm{pc}^{-3}$. We adopt the values from \cite{2016barros} of the surface density near the Sun as $\Sigma_0 = 30.2 \;M_\odot \textrm{pc}^{-2}$, the half thickness of the thin disk as $h = 205 \textrm{pc}$, the disk radius of $R_d = 2.12$ kpc, and the central hole in the thin disk at $R_\textrm{ch} = 2.07$ kpc\footnote{At our 150 pc radius, the first two terms of Equation \ref{eq:mass} contribute only one percent to the total, but we include the full equation here for completeness.}. $R_0$ is the distance from the Sun to the Galactic Center, $R_0 = 8.1$ kpc.

Finally, we express the values of $z$ and $R$ in observable quantities:
\begin{gather}
    z = r\sin b; R^2 = (r\cos b)^2 = R_0^2 - 2 R_0 (r\cos b) \cos \ell
    \label{eq:coords}
\end{gather}

We note that the Milky Way model by \cite{2016barros} implies thin disk stellar masses of $2.49\times10^{10} M_\odot$ and $4.02\times10^{5} M_\odot$ for the entire Galaxy and within 150 pc, respectively.

\subsubsection{Final Values}

After having done the above calculations, we compute three space densities:
\begin{gather}
    \rho_M = \sum_i \frac{1}{\delta M_{i}}; \rho_{N} =\rho_* \sum_i \frac{1}{\delta M_{i}}; \rho_L = \sum_i \frac{L_{X,i}}{\delta M_{i}}
    \label{eq:xlf}
\end{gather}
where $\rho_*$ is the midplane stellar mass density\footnote{We use the midplane value of $0.0736 M_\odot\textrm{pc}^{-3} $ from \cite{2016barros}, whereas \cite{2022suleimanov} convert to the local stellar mass density at the solar height of 20.8 pc above the midplane.} (in $M_\odot \textrm{pc}^{-3}$), and $L_{X,i}$ is the X-ray luminosity for the $i^\textrm{th}$ system. Because we calculate the number density at the midplane, we report the quantity as $\rho_{N,0}$. The sum is then carried out for all systems in the sample. If a sample out to greater distances is indeed complete and assumptions satisfied, repeating this calculation should converge to the same space densities. We calculate space densities for AM CVns separately, since they have different formation channels than CVs  (i.e. space densities calculated for CVs have AM CVns removed).

We assume Poisson uncertainties (adopting 68\% confidence intervals) in obtaining the total number of systems, and report all space densities in Table \ref{tab:xlf}, noting that our uncertainties can accommodate $N_\textrm{CV} = 20\pm4$ CVs and $N_\textrm{AM CVn} = 3\pm2$ AM CVns within 150 pc (in half of the sky). We note that the above calculation effectively assumes that CVs trace the Galactic thin disk --- and therefore a CV scale height of 205 pc. We explore the effect of adopting different scale heights in Appendix \ref{sec:h_cv}. 

\begin{figure*}
    \centering
    \includegraphics[width=0.31\textwidth]{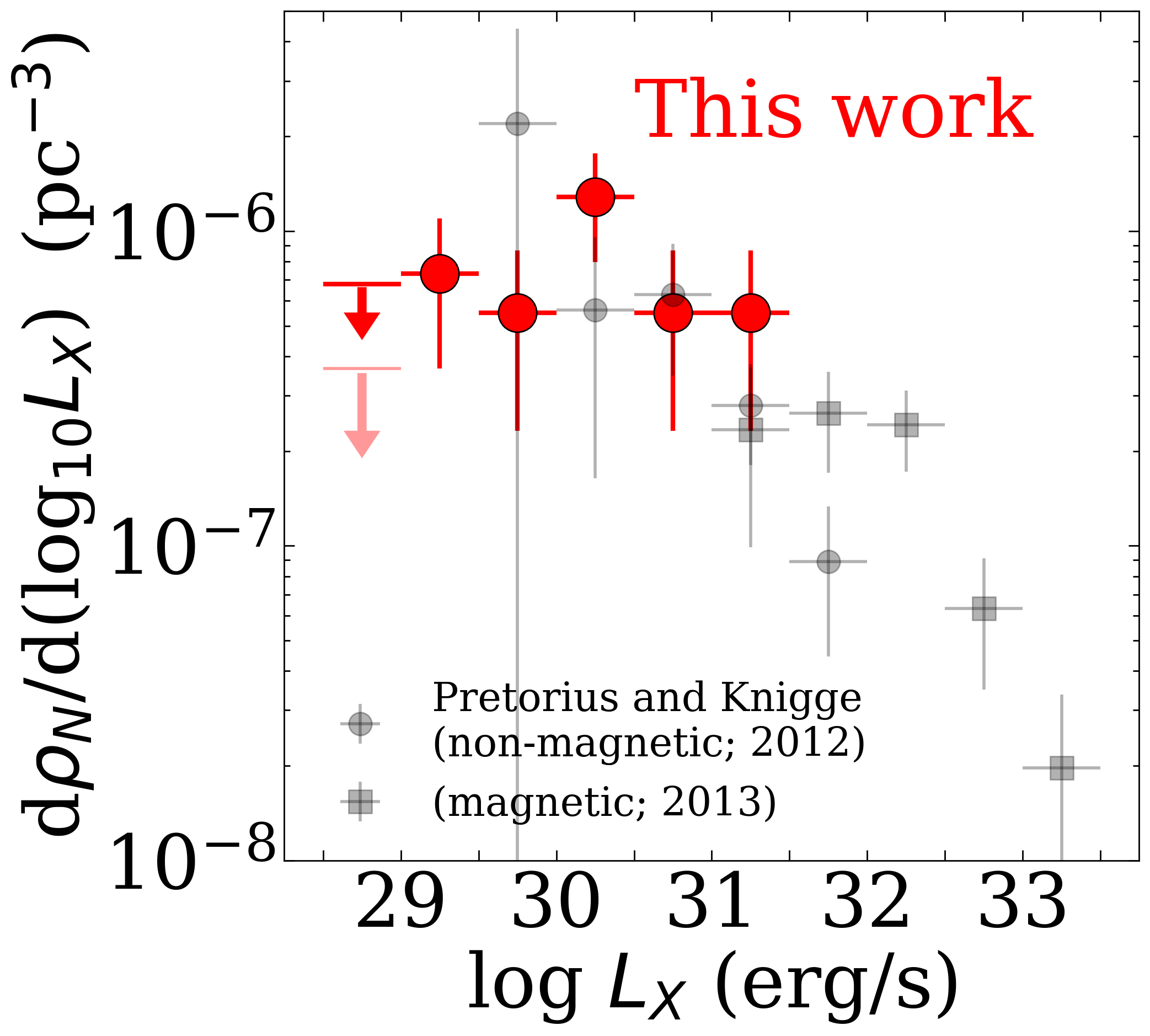}\includegraphics[width=0.31\textwidth]{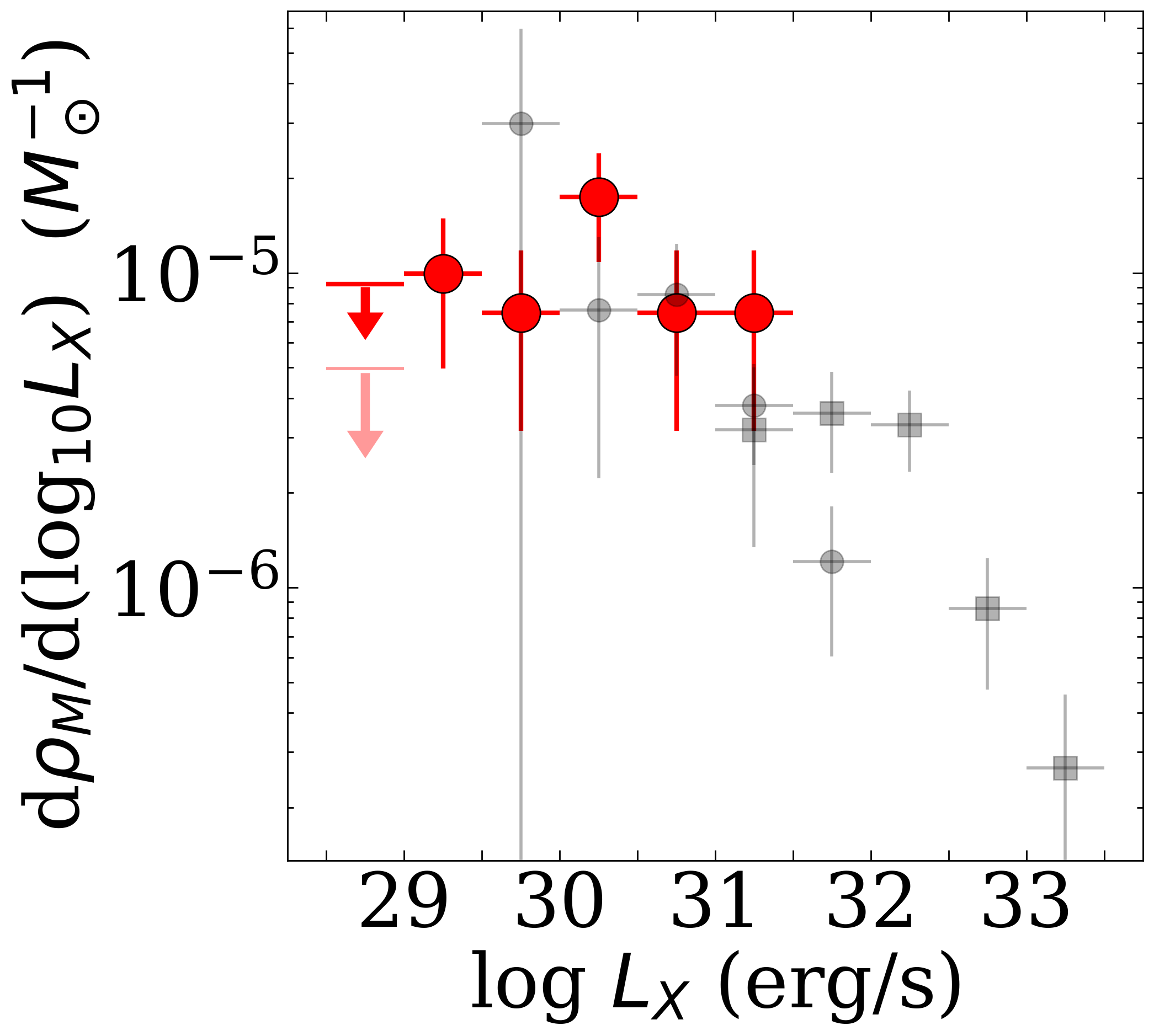}\includegraphics[width=0.335\textwidth]{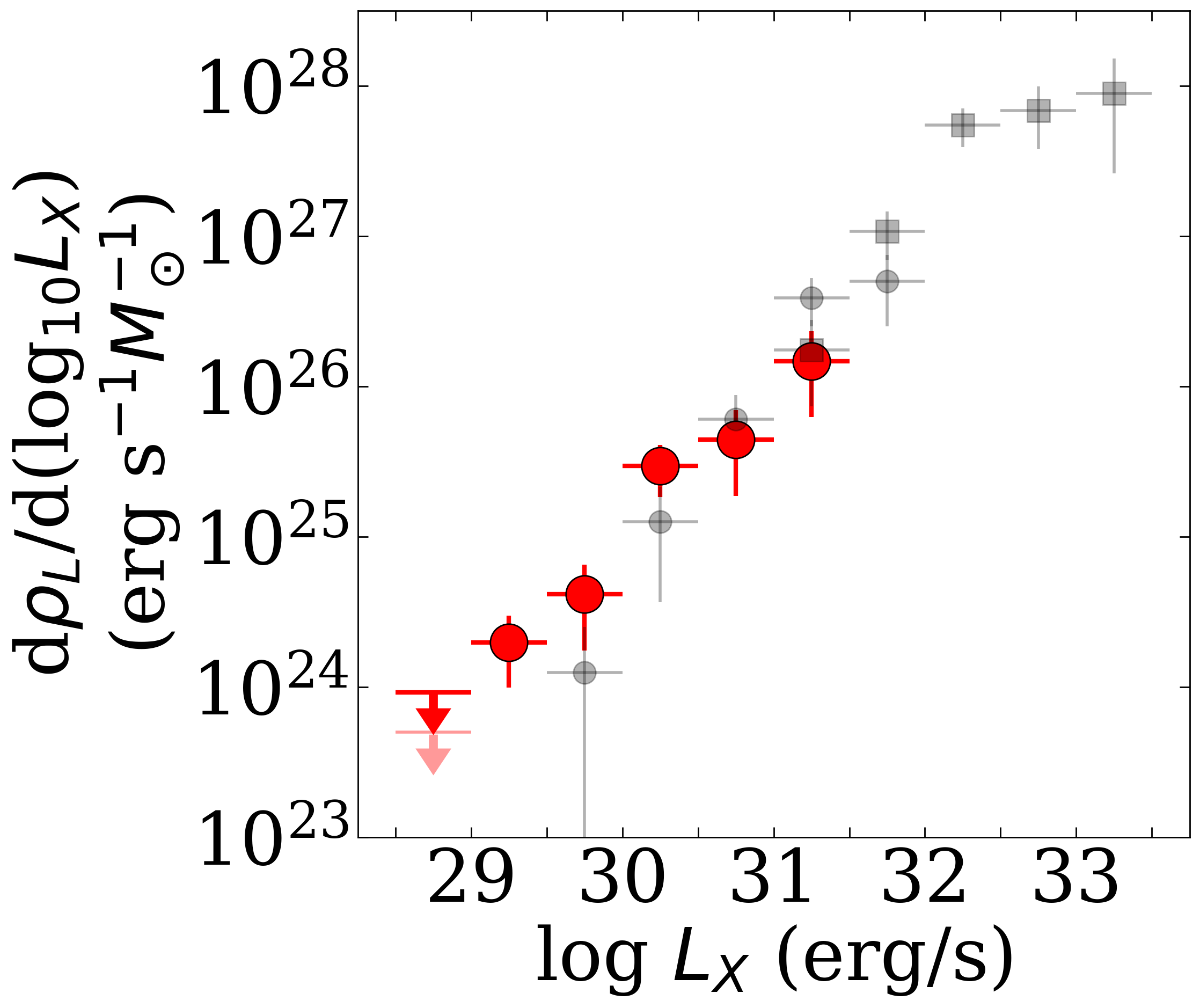}
    \caption{X-ray luminosity functions of the CVs in our 150 pc sample (red points), with AM CVns excluded. A clear flattening is seen around $L_X \sim 10^{30} \textrm{erg s}^{-1}$ in all number and mass luminosity functions (left and 
    center panels), demonstrating that we are probing the lowest $L_X$ end of CVs. A significant number of low $L_X$ systems lead to smaller error bars compared to previous work (gray points). Low $L_X$ ($\lesssim10^{30} \textrm{erg s}^{-1}$) CVs dominate in number over high $L_X$ ($\gtrsim10^{30} \textrm{erg s}^{-1}$) systems (left and center panels), while the right panel shows that the few high $L_X$ systems dominate the total luminosity budget of the population. }
    \label{fig:xlf}
\end{figure*}

We present X-ray luminosity functions in $L_X$, in Figure \ref{fig:xlf}. We only include CVs, not AM CVns. As a comparison, we also plot previous X-ray luminosity functions of non-magnetic CVs \citep{2012pretorius} and magnetic CVs \citep{2013pretorius}. We have chosen these two works since the ROSAT 0.1--2.4 keV energy range is most similar to the 0.2--2.3 eROSITA energy range; other studies have focused on CVs in hard X-rays \citep[e.g.][]{2008Revnivtsev, 2022suleimanov}. We adopt the values of effective volume and X-ray luminosity reported in those papers, to provide a fair comparison, and compute X-ray luminosity functions using Equation \ref{eq:xlf}. We note that those works use pre-\textit{Gaia} distances and compute effective volumes using the ``1/$V_\textrm{max}$" method \citep{1993tinney}. We just use this as a \textit{comparison} sample, mainly to highlight our sensitivity to low $L_X$ systems.

Figure \ref{fig:xlf} shows, for the first time, a ``flattening" in the X-ray luminosity function of CVs --- i.e. they no longer keep increasing as one goes to lower values of $L_X$ as they have been in every previous plot of CV X-ray luminosity functions \citep[e.g.][]{2008Revnivtsev, 2012pretorius, 2013pretorius, 2018schwope, 2022suleimanov}. We note that the X-ray luminosity function presented by \cite{2006sazonov} revealed a flattening and a turn over, though only when both CVs and active binary stars were included. The number and mass density panels show that low X-ray luminosity ($L_X\sim10^{30} \textrm{erg s}^{-1}$) make up the dominant population of CVs. The luminosity density panel shows that our sample of these low-$L_X$ systems increases their contribution to the total $L_X$ budget of CVs by a factor of a few, while the high X-ray luminosity ($L_X \sim 10^{32} - 10^{33} \textrm{ erg s}^{-1}$) make up the dominant X-ray output of CVs. Thus, in any given population of CVs, it will be the low $L_X$ systems that will be greatest in number, but it will be the high $L_X$ systems that dominate the overall X-ray luminosity. We compute upper limits in the $10^{28.5}< L_X<10^{29} \textrm{erg s}^{-1}$ are assuming one system within 94 pc (dark red bar) or two systems within 150 pc (light red bar). The former assumes there is at most one system out to the distance where eRASS1 is complete for $L_X = 10^{28.5} \textrm{erg s}^{-1}$, which is 94 pc. The latter assumes that the two systems in the \cite{2020pala} sample that we do not detect in eRASS1 are the only two systems within 150 pc that have $L_X < 10^{29} \textrm{erg s}^{-1}$. 

As a final note, we emphasize that the space densities and luminosity functions presented here are representative of the solar neighborhood, but likely do not extend to other Galactic environments such as regions of the thin disk closer to the Galactic Center, the nuclear star cluster, globular clusters, etc. For that reason, we cannot reliably estimate the total number of CVs in the Galactic Center and/or Galactic Ridge. Other works, for example, \cite{2022suleimanov}, were able to do so, since the hard X-ray flux limited survey used in that study was sensitive to more distant systems. In our case,  changes in the initial mass function, star formation rate, and binary fraction in different Galactic environments would likely lead to different number densities of CVs (per volume or per solar mass) as well as a different distribution of $L_X$ compared to our local sample. 

\begin{table*}
    \centering
    \begin{tabular}{l||l|l|l|l|l}
        \parbox[c][1.5cm][c]{4.5cm}{\centering Sample} &
        \parbox[c][1.5cm][c]{2.2cm}{\centering $\langle L_X\rangle$ \\($\textrm{erg s}^{-1}$)} &
        \parbox[c][1.5cm][c]{2.0cm}{\centering $\langle \log L_X\rangle$} &
        \parbox[c][1.5cm][c]{2.2cm}{\centering $\rho_{N,0}$ \\($\textrm{pc}^{-3}$)} &
        \parbox[c][1.5cm][c]{2.2cm}{\centering $\rho_M$ \\($M_\odot\textrm{pc}^{-3}$)} &
        \parbox[c][1.5cm][c]{2.2cm}{\centering $\rho_{L_X}$ \\($\textrm{erg s}^{-1} \textrm{pc}^{-3}$)} \\
        \hline
         150 pc CVs (no AM CVns; 20)&$4.6\pm1.5\times 10^{30}$& $30.19\pm 0.15$& $3.7\pm0.7\times10^{-6}$&$5.0\pm1.0\times10^{-5}$&$2.3\pm0.4\times10^{26}$\\\hline
         150 pc AM CVns (3) &$8.6\pm3.5\times 10^{29}$& $29.86\pm 0.19$& $5.5\pm3.7\times10^{-7}$&$7.5\pm5.0\times10^{-6}$&$6.4\pm4.3\times10^{24}$\\
         150 pc Non-magnetic CVs (13)&$3.6\pm1.6\times 10^{30}$& $30.19\pm 0.16$& $2.4\pm0.7\times10^{-6}$&$3.2\pm1.0\times10^{-5}$&$1.2\pm0.4\times10^{26}$\\
         150 pc Magnetic CVs (7)&$6.3\pm3.3\times 10^{30}$& $30.19\pm 0.34$& $1.3\pm0.5\times10^{-6}$&$1.7\pm0.7\times10^{-5}$&$1.1\pm0.5\times10^{26}$\\
         150 pc Period Bouncers (5) &$3.4\pm1.3\times 10^{29}$& $29.42\pm 0.14$& $9.2\pm2.7\times10^{-7}$&$1.2\pm0.5\times10^{-5}$&$4.2\pm2.7\times10^{24}$\\\hline
         VSX CVs + ROSAT&$1.1\pm0.3\times 10^{32}$& $31.42\pm0.06$& -&-&-\\
         VSX CVs + eROSITA&$4.8\pm1.3\times 10^{31}$& $30.97\pm0.02$& -&-&-\\
         150 pc CVs (+ AM CVns; 23)&$4.0\pm1.4\times 10^{30}$& $30.15\pm 0.14$& -&-&-   
    \end{tabular}
    \caption{Average X-ray luminosities and space densities for all CVs, AM CVns, and CV subtypes in our sample. Our average $L_X$ value of CVs is at least a factor of 10 lower than that computed from samples of optically identified systems (VSX) with X-ray counterparts (ROSAT and eROSITA). Magnetic CVs and period bouncers make up 35\% and 25\% of CVs in our sample, respectively.}
    \label{tab:xlf}
\end{table*}

\subsection{New CVs Within 150 pc}
\label{sec:new_systems}

We discovered two new systems within 150 pc in the Western Galactic Hemisphere: 1eRASS J101328.7-202848, a new AM CVn which is the third closest known AM CVn (and ultracompact binary) to Earth at 140 pc, and 1eRASS J054726.9+132649, a new magnetic period bouncer. 

\subsubsection{1eRASS J101328.7-202848: A New AM CVn}

1eRASS J101328.7-202848 (\textit{Gaia} DR3 ID: 5668919328172331904) is a long-period ($P_\textrm{orb}\gtrsim 50$ min) AM CVn located 139$\pm3$ pc away. Follow-up spectroscopy and/or high cadence photometry are needed to confirm the orbital period. However, the location of this system near the WD track (see Figure 13 in \cite{2023rodriguez_amcvn}) implies that it is a long-period ($P_\textrm{orb}\gtrsim 50$ min) system. In addition, the lack of outbursts in the five year long archive of ZTF data is indicative of a low mass transfer rate, and therefore a long orbital period. Previous works have shown that $P_\textrm{orb}\gtrsim 50$ min systems tend to not show outbursts on the timescales of current photometric surveys \citep[e.g.][]{2018ramsay, 2021duffy, 2022vanroestel, 2023rodriguez_amcvn}.

\begin{figure*}
    \centering
    \includegraphics[width=\textwidth]{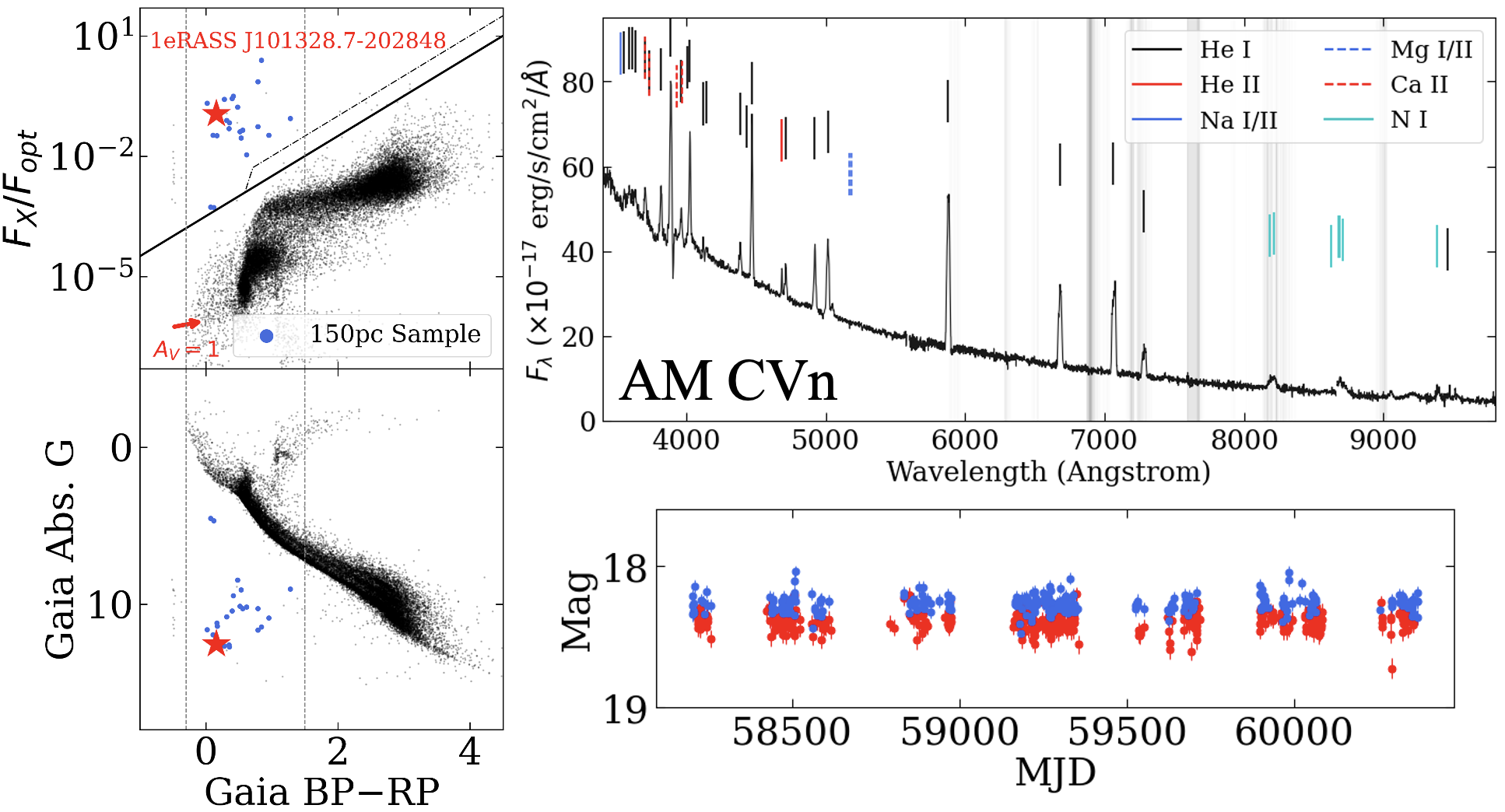}
    \caption{1eRASS J101328.7-202848 is a newly-discovered AM CVn, selected thanks to the X-ray Main Sequence (upper left). The optical spectrum (upper right) shows prominent He I/II emission lines, an absence of H lines, and the presence of metals resulting from the donor polluting the WD. Gray lines indicate telluric features. It appears near the WD track on the HR diagram (lower left), and shows no outbursts in six years of ZTF data (lower right). Those properties suggest it is a long-period ($P_\textrm{orb}\gtrsim 50$ min) AM CVn.}
    \label{fig:amcvn}
\end{figure*}

The spectrum of 1eRASS J101328.7-202848 (Figure \ref{fig:amcvn}) shows a blue continuum with prominent He I lines, some of which are double-peaked (e.g. He I 7065, 7281), and even single-peaked He II 4686. There are no H lines. There are also Ca II and Mg I/II lines in absorption and N I lines in emission, as have been seen in other long-period AM CVns and attributed to WD pollution by the donor \citep[e.g.][]{2022vanroestel, 2023rodriguez_amcvn}. We will report on phase-resolved spectroscopy and high-cadence photometry of this object in a future study.

Figure \ref{fig:amcvn} reveals the location of 1eRASS J101328.7-202848 in the X-ray Main Sequence, clearly above the cut to select accreting compact object binaries. We also show its location in the \textit{Gaia} HR diagram and its quiescent behavior in the five year long ZTF archive. We obtained a single 600 sec optical spectrum with the Low Resolution Imaging Spectrometer \citep[LRIS;][]{1995lris} on the Keck I telescope on 31 March 2024. The 400/3400 grism was used and 400/8500 grating with the D55 dichroic. The seeing was $<$1", and the 1" long slit was used, leading to minimal slit losses. Data were wavelength calibrated with internal lamps, flat fielded, and cleaned for cosmic rays using \texttt{lpipe}, a pipeline for LRIS optimized for long slit spectroscopy \citep{2019perley_lpipe}.

\subsubsection{1eRASS J054726.9+132649: A New Magnetic Period Bouncer}
\label{sec:new_pb}

\begin{figure*}
    \centering
    \includegraphics[width=\textwidth]{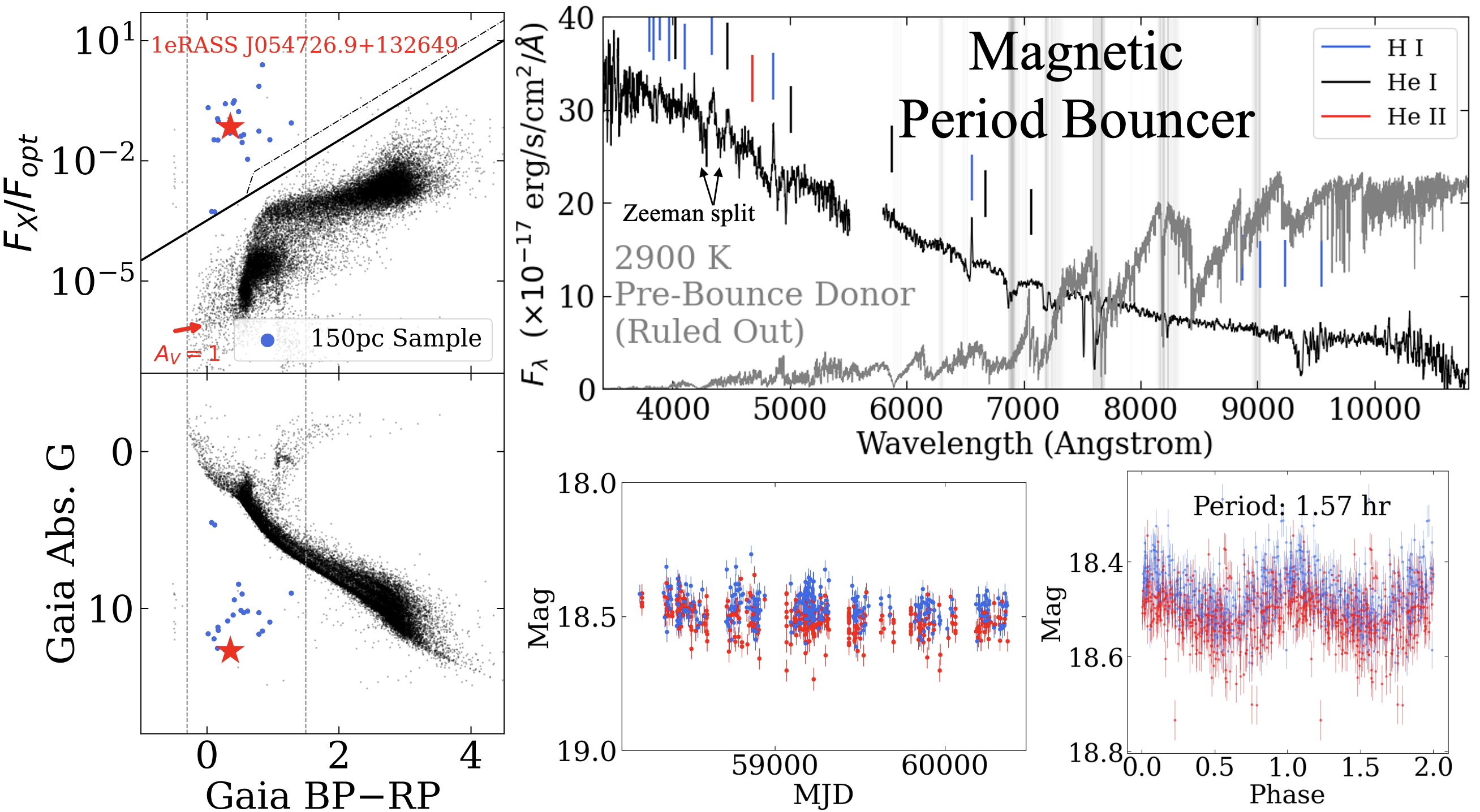}
    \caption{1eRASS J054726.9+132649 is a newly-discovered magnetic period bouncer, selected thanks to the X-ray Main Sequence (upper left). It is located near the WD track (bottom left), and shows no outbursts in six years of data (bottom center), indicative of low mass transfer rates. ZTF optical photometry ($r$-band in red, $g$-band in blue) reveals a 1.57 hr orbital period (bottom right). The optical spectrum (upper right) shows that a 2900 K donor star predicted by CV evolutionary tracks (see Figure \ref{fig:pb_proof}) at this orbital period is not seen, thus confirming this system as a period bouncer. Balmer emission lines show clear Zeeman splitting (particularly H$\beta$ and H$\gamma$), which reveal the magnetic nature of the WD. }
    \label{fig:pb}
\end{figure*}

1eRASS J054726.9+132649 (\textit{Gaia} DR3 ID: 3346524412647596032) is a magnetic period bouncer with a 93.95-min (1.565 hr) orbital period, 132$\pm$3 pc away. Its magnetic classification comes from the obvious Zeeman splitting of the H$\beta$ and H$\gamma$ lines in the spectrum (Figure \ref{fig:pb}). We present the ZTF $r$ and $g$ light curves of this system in Figure \ref{fig:pb}, which reveal its clear periodicity. We found this period based on a Lomb-Scargle periodogram using the \texttt{gatspy} tool \citep{2016gatspy}. We searched for periods between 5 minutes and 10 days, with a linear frequency grid oversampled by a factor of 10 and found no other significant periods. The ephemeris we report based on the ZTF data is $t_0$ (BJD)= 2458381.3339(2) + $T\times$ 0.06524(1). 

This system bears a remarkable similarity to SDSS J125044.42+154957.3, a polar also in our 150 pc sample. SDSS J125044.42+154957.3 has an 86.3 min orbital period and its white dwarf has a magnetic field strength of 21 MG. SDSS J125044.42+154957.3 has a solid classification based on phase-resolved spectroscopy as a polar near the CV period minimum \citep{2012breedt}, but has been recently labeled as a period bouncer due to its low X-ray luminosity and late M (near L) spectral type \citep{2023munozgiraldo}.

In Figure \ref{fig:pb}, we show the location of 1eRASS J054726.9+132649 in the X-ray Main Sequence, clearly above the cut to select accreting compact object binaries. We also show its location in the \textit{Gaia} HR diagram and its quiescent behavior in the five year long ZTF archive. It would not be detectable in a search for outbursting systems, but its periodicity would clearly be revealed in a search for systems showing such behavior. 

Now, we justify why 1eRASS J054726.9+132649 1) is a true polar with a Roche lobe filling donor instead of a wind-accreting, low accretion rate polar, and 2) has evolved past the period minimum. To address the first point, we note the arguments put forth by \cite{2012breedt}: Roche lobe filling CVs should show a hotter WD than those in wind-accreting systems. High states of accretion in Roche lobe filling systems regulate the temperature of the WD \citep{2003townsley} and keep it higher than those in wind accreting (or detached) systems. Moreover, the orbital periods of wind accreting systems is generally $\gtrsim$2 hr, since some of these could be pre-polars that have not yet filled their Roche lobes \citep{2012breedt, 2022hakala}. The spectrum of 1eRASS J054726.9+132649 in Figure \ref{fig:pb} shows a blue continuum characteristic of an accretion-heated WD, but quantitative proof of this is clear from the \textit{Gaia} BP--RP color of this system being nearly equal to that of SDSS J125044.42+154957.3. In Figure \ref{fig:all}, they nearly overlap on the HR diagram and are remarkably close together in the X-ray Main Sequence due to their similar values of X-ray luminosity. We thus conclude that 1eRASS J054726.9+132649 is a true polar, with a Roche lobe filling donor. 

\begin{figure}
    \centering
    \includegraphics[width=0.45\textwidth]{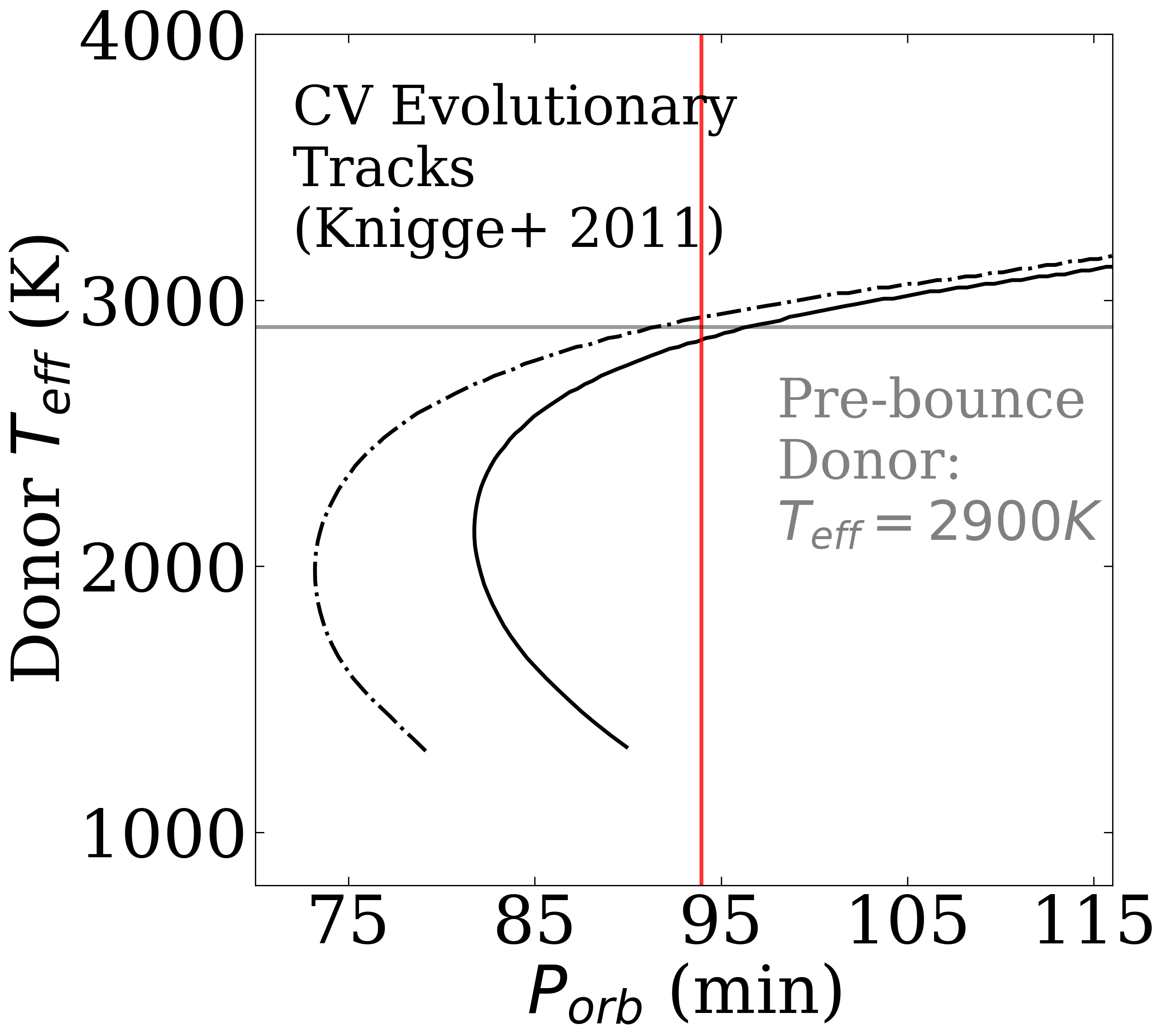}
    \caption{The orbital period of 1eRASS J054726.9+132649 is shown in red, and CV ``optimal" and ``standard" evolutionary tracks are shown by the solid and  dot-dashed lines, respectively. If this were a pre-bounce system, its donor $T_\textrm{eff}$ would be 2900K.  Since we do not see this in the spectrum in Figure \ref{fig:pb}, we determine it must  be a period bouncer. }
    \label{fig:pb_proof}
\end{figure}

To address the second point, we include Figure \ref{fig:pb_proof}. At the orbital period of 1eRASS J054726.9+132649, if the system evolves according to the evolutionary tracks of \cite{2011knigge}, then the donor star should have an effective temperature of $T_\textrm{eff}\approx$ 2900 K in either the ``standard" or ``optimal" track. In Figure \ref{fig:pb}, we plot a BT-DUSTY model atmosphere \citep{2011dusty} of that effective temperature, with the corresponding radius in the \cite{2011knigge} tracks of $0.15 R_\odot$, at the \textit{Gaia} median distance of 132 pc. In Figure \ref{fig:pb}, the observed spectrum does not show evidence of such a donor, particularly redward of $\approx$ 7000 \AA. The presence of an accretion stream would only increase the flux level, so we take this as conclusive evidence that a donor star with the temperature and radius that we would expect at the orbital period of 1eRASS J054726.9+132649 is not present. This means that if 1eRASS J054726.9+132649 follows standard CV evolutionary tracks, it must be a period bouncer. Similar arguments have been put forth for other period bouncers in the literature: QZ Lib and SRGe J041130.3+685350 \citep{2018pala, 2024galiullin}. Finally, we show in Figure \ref{fig:all} that its low X-ray luminosity is consistent with that of other period bouncers (both magnetic and non-magnetic) at similar orbital periods. 

We obtained a single 900 sec optical spectrum with the Double Spectrograph \citep[LRIS;][]{1995lris} on the Hale 200-inch telescope at Palomar Observatory on 21 March 2024. The 600/4000 grism was used and 316/7600 grating with the D55 dichroic. The seeing was $<$2", and the 1.5" long slit was used, leading to minimal slit losses. Data were wavelength calibrated with internal lamps, flat fielded, and cleaned for cosmic rays using \texttt{DBSP-DRP}\footnote{\url{https://dbsp-drp.readthedocs.io/en/stable/index.html}}, a Python-based pipeline optimized for DBSP built on the more general \texttt{PypeIt} pipeline \citep{2020pypeit}.

\section{Discussion}
\label{sec:discussion}
\subsection{Sample Completeness and No Very Low $L_X$ CVs}
\begin{figure}
    \centering
    \includegraphics[width=0.47\textwidth]{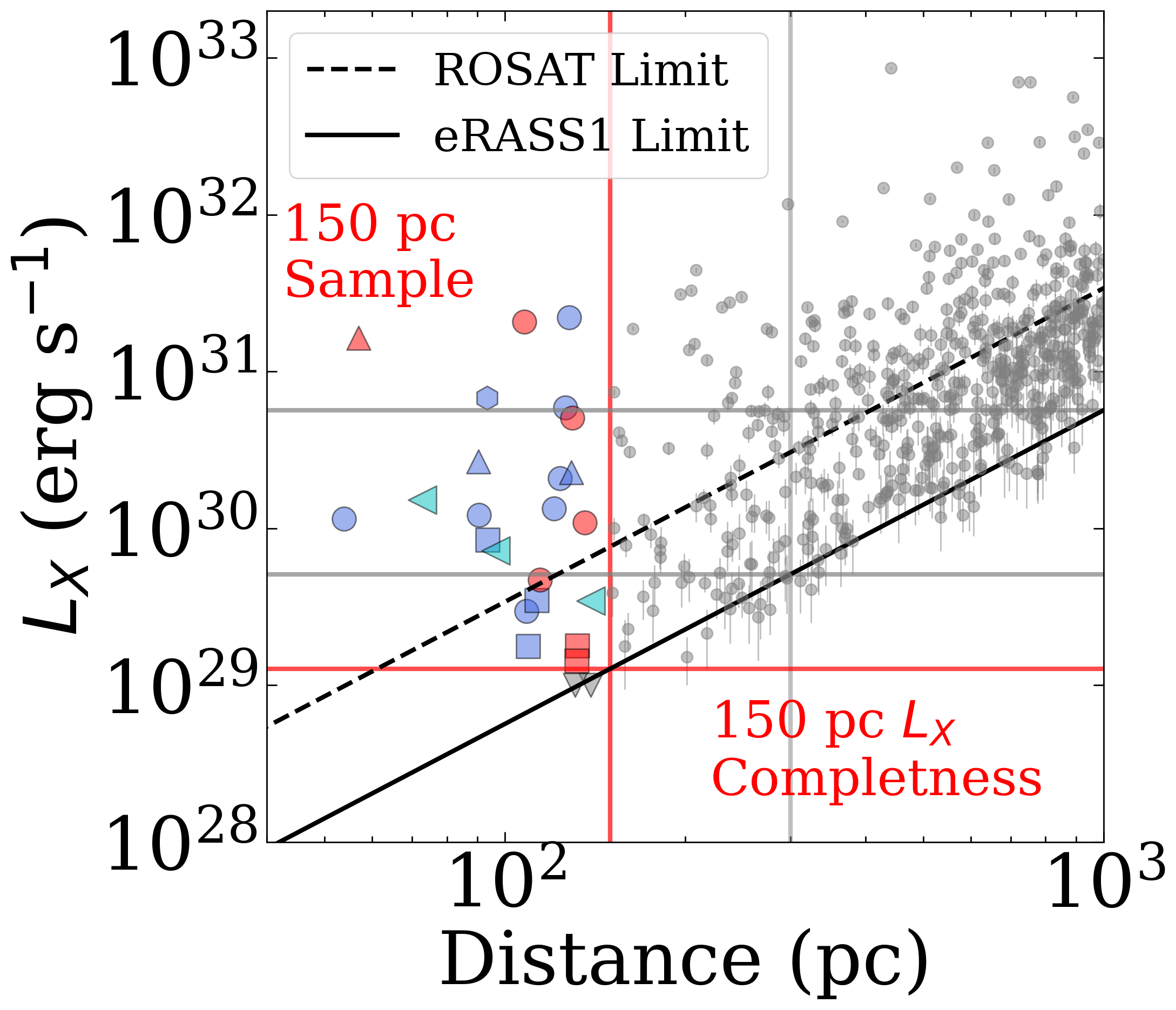}
    \caption{All CVs in the 150 pc sample are shown using the same colors/markers as in Figure \ref{fig:all}, with the two systems in the \cite{2020pala} sample not in eRASS1 shown as downward pointing gray triangles. All CV candidates in our 1000 pc sample are shown as gray circles. Gray lines indicate the $L_X$ completeness limits at 300 pc and 1000 pc, demonstrating that our 150 pc sample at the depth of eRASS1 is the only combination of distance/survey to obtain a complete CV catalog. }
    \label{fig:limits}
\end{figure}

In Figure \ref{fig:limits}, we plot the SRG/eROSITA eRASS1 0.2--2.3 keV X-ray luminosity, $L_X$ as a function of distance, $d$, of all systems in the 1000 pc sample. We emphasize that a CV survey at the depth of eRASS1 out to a distance of 150 pc is the \textit{only} current combination of survey/distance that can be sensitive to all CVs. The flux limit of the ROSAT 2RXS catalog renders it complete down to only $L_X\sim 8\times 10^{29}\textrm{ erg s}^{-1}$, at 150 pc, missing nearly half of all objects in our sample. Crucially, two of the three AM CVns and four of the five period bouncers would be missed, both of which are arguably the most poorly-understood CV subtypes and critical to our understanding of CV evolution. 

Figure \ref{fig:limits} also demonstrates that low $L_X$ CVs do not make up a significant fraction of all systems. At $\sim$100 pc, we reach an X-ray completeness of $L_X = 5\times 10^{28} \textrm{erg s}^{-1}$. This means that if there were a dominant population of CVs at such low X-ray luminosities, which was not able to be ruled out by previous CV surveys \citep[e,g,][]{2012pretorius}, we would have certainly detected it. 

At the luminosity limit of eRASS1 out to 300 pc, nearly a third of the CVs in our 150 pc sample would be missed, including four of the five period bouncers. At the $L_X$ limit at 1000 pc, a full 80\% of CVs in our 150 pc sample would be missed, demonstrating that complete samples of CVs can only be carried out to nearby distances.

\begin{figure*}
    \centering
    \includegraphics[width=0.9\textwidth]{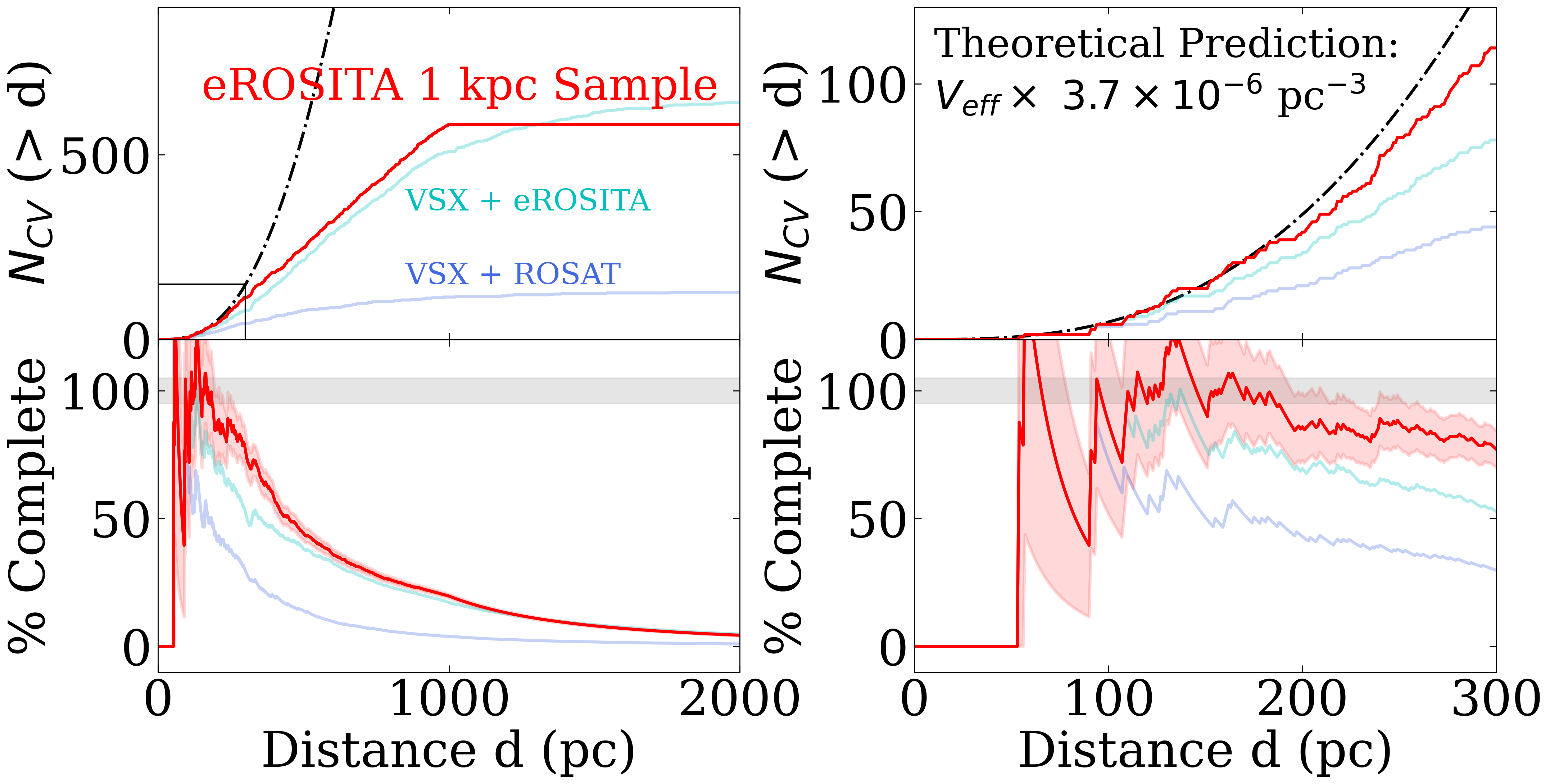}
    \caption{The cumulative distribution of CVs is plotted as a function of distance (upper panels) for our 1000 pc sample (red), and VSX samples. The predicted number of systems given our estimate of $\rho_{N,0}$ (black) indicates that both VSX samples are incomplete, even at $d\lesssim$ 100 pc. The percentage of the difference between the predicted and observed cumulative distributions is shown in the lower panels. At distances beyond $\approx$ 180 pc, even our eROSITA + \textit{Gaia} sample starts to become incomplete.} 
    \label{fig:complete}
\end{figure*}

In order to quantify the completeness of our sample, we compare the cumulative number of CVs (as a function of distance) to the effective volume probed by our calculated number density, in a similar manner to \cite{2021el-badry}. In upper panels of Figure \ref{fig:complete}, we show the cumulative number of CVs in our 1000 pc sample, as well as the VSX + ROSAT and VSX + eROSITA comparison samples. There are more CVs in our 1000 pc sample compared to either of the VSX sample, signalling that we have likely constructed the largest sample of X-ray selected CV candidates to date. The cumulative number of our 1000 pc is overall higher, meaning that the primarily optically-constructed VSX catalog is incomplete at all distances. 

We also overplot the effective volume, multiplied by the space density we derive for CVs. This is effectively the theoretical total number we expect as a function of distance, which agrees with the cumulative number of systems in our sample up until a distance of $\approx$ 180 pc. Beyond that distance, the completeness plummets. Neither of the VSX samples are complete at any distance. We demonstrate this graphically in the bottom panels of Figure \ref{fig:complete}, plotting the percent difference between the theoretical and observed cumulative numbers of each sample, assuming Poisson counting errors.  

\subsection{Comparison to Previous Estimates of Average $L_X$}
\label{sec:prev_studies}

\subsubsection{CVs}
One of the major findings of this work is that the mean X-ray luminosity of most CVs is $\langle L_X\rangle \sim 10^{30}\textrm{erg s}^{-1}$ in the 0.2--2.3 keV range. We also showed that the naive construction of an X-ray luminosity distribution based on X-ray detections of previously known, \textit{primarily optically-identified} systems leads to an overestimate of this, with $\langle L_X\rangle \sim 10^{31}- 10^{32}\textrm{erg s}^{-1}$. This implies that our 1000 pc sample is likely rich with true CVs that have been missed by previous X-ray surveys. 

An important exception to this, of course, are targeted X-ray observations (with \textit{Swift}, \textit{XMM-Newton}, \textit{Chandra}, etc.) of systems identified by other means. An example of such a study was conducted by \cite{2013reis}. That study mainly targeted low accretion rate (WZ Sge-type) CVs discovered by SDSS, the majority of which had orbital periods near the period minimum $P_\textrm{orb} < 90$ min. \cite{2013reis} obtained \textit{Swift}/XRT follow-up of $\sim$20 systems, and found $\langle \log L_X (\textrm{erg s}^{-1}) \rangle = 29.78$, and $ \langle L_X \rangle = 8 \times 10^{29}\textrm{erg s}^{-1}$ in the 0.5--10 keV range. In Figure \ref{fig:all}, we show that many of our systems in a similar period range also have similarly low values of $L_X$. However, the systems used to make up the \cite{2013reis} sample were primarily non-magnetic CVs, not selected homogeneously, so a one-to-one comparison to our result is not possible.

Interestingly, \cite{2012pretorius} note that applying a classic model of CV X-ray emission \citep{1985patterson} to a synthetic population that (correctly) predicts the majority of CVs should be low accretion rate systems below the period gap \citep{1993kolb}, predicts the majority of systems should have X-ray luminosities between a few $\times 10^{29}\textrm{erg s}^{-1}$ and a few $\times 10^{30}\textrm{erg s}^{-1}$, which is similar to what we have found here. In short, while various earlier studies have postulated that there should be a large population of low $L_X$ CVs, a secure discovery of this has not been possible until now. 

While this work was in preparation, \cite{2024schwope_new} independently reported on SRG/eROSITA detections of CVs from known catalogs, and calculated $\langle\log L_X\rangle$ for each catalog. However, that work only reported on detections of previously known CVs, and did not conduct a search for new systems as we have done here, so those average $L_X$ values cannot be taken to be representative of CVs. \cite{2024schwope_new} found $\langle\log L_X\rangle = 30.34$ for the systems found in the \cite{2020pala} sample, though, as we have shown here, that sample is missing two systems within 150 pc.

\subsubsection{AM CVns}
The largest catalog of X-ray detections of AM CVns was compiled by \cite{2023begari}, who searched the \textit{XMM-Newton}, \textit{Chandra} and \textit{Swift} archives for X-ray counterparts to known AM CVns. They showed a clear trend of decreasing X-ray luminosity as a function of orbital period. Our average $L_X$ of $8.6 \pm 3.5 \times 10^{29} \textrm{ erg s}^{-1}$ is consistent with that of their longest period systems. Crucially, \cite{2006ramsay} found that as AM CVns evolve to longer orbital periods, a large fraction luminosity is emitted in the ultraviolet, dominating over the X-ray. At this point, larger samples of AM CVns, particularly X-ray bright, short period systems, are needed to determine the average luminosity density, and if, as is the case with CVs, that the few X-ray bright systems dominate the total luminosity output over the dominant population of low $L_X$ systems. 

\subsection{Synergy with Flux-Limited Surveys}

Volume-limited surveys are by no means the only way to discover CVs. X-ray flux-limited and optical magnitude-limited surveys are the only way to discover intrinsically rare, yet scientifically interesting CVs that probe short evolutionary stages. In Figure \ref{fig:lum_dist}, it is clear that CVs with $L_X\gtrsim10^{32} \textrm{ erg s}^{-1}$ are only found at $d \gtrsim$ 300 pc. Indeed, many of these systems are long-period IPs, which are abundant in (hard) X-ray surveys and form a crucial part of understanding of magnetism in CVs \citep{2019suleimanov}. 

As mentioned previously, this is especially the case for AM CVns, where the moderate period $P_\textrm{orb}\lesssim$ 50 min systems, and particularly the short period $P_\textrm{orb}\lesssim$ 20 min systems, are only found at distances beyond a few hundred pc \citep{2018ramsay}. These systems tend to have higher $L_X$ values \citep{2023begari}, and many have indeed been discovered through X-ray surveys.

The point of this work (and volume-limited surveys in general) is not to be sensitive to \textit{all} CVs, but to conduct a systematic survey that detects all nearby CVs, which should be representative of the population in all regions of the Milky Way that resemble the solar neighborhood. In Figure \ref{fig:lum_dist}, it is clear that low $L_X$ systems, like long-period AM CVns and period bouncers, cannot be detected beyond 150--300 pc given the flux limit of eRASS1. These CV subtypes, once believed to be rare, are actually abundant in the Milky Way, but will remain invisible until the advent of deeper all-sky X-ray and optical surveys such as future releases from SRG/eROSITA and the upcoming Rubin Observatory Legacy Survey of Space and Time (LSST).

\subsection{Comparison to \cite{2020pala}}

We wish to emphasize the importance of the 150 pc survey carried out by \cite{2020pala} in this work. 21 of the 23 CVs in our sample were already compiled by \cite{2020pala}, spectroscopically verified either by that work or before, and had orbital periods determined. We acknowledge that most of the conclusions we reach here echo those of \cite{2020pala}, though 1) all X-ray analysis presented here is new, and 2) we have made use of a novel selection tool, the X-ray Main Sequence, which has single-handedly recovered virtually all systems in the \cite{2020pala} sample, regardless of subtype. Crucially, the latter allows us to derive space densities without adopting a completeness correction, which is sensitive to the targeting strategy of SDSS \citep{2020pala, 2023inight}.

We now speculate as to why we may have missed two systems in the \cite{2020pala} sample within 150 pc in the Western Galactic Hemisphere: Gaia J154008.28–392917.6 ($d=139$ pc) and V379 Tel  ($d=131$ pc). Visual inspection of the eRASS1 images confirms they are not detected. Given the flux limit of eRASS1 and the well-measured distance, it is possible that Gaia J154008.28–392917.6 has an X-ray luminosity of $L_X\lesssim1.0\times10^{29} \textrm{erg s}^{-1}$. \cite{2020pala} showed that this system is a WZ Sge-type low accretion rate CV that has shown no outbursts, so the possibility of a low X-ray luminosity is likely. Such low values of $L_X$, though rare, have been seen in CVs --- as mentioned previously, one of the twenty CVs in the sample of \cite{2013reis} indeed has $L_X<1.3\times 10^{29} \textrm{erg s}^{-1}$. In the case of V379 Tel, there are two possible scenarios. Firstly, since this system is an \textit{eclipsing} polar \citep{2005v379}, is possible that it was missed by the single eROSITA scan of this part of the sky. Using the eROSITA upper limit calculator\footnote{\url{https://erosita.mpe.mpg.de/dr1/erodat/upperlimit/single/}}, the exposure time at this position was 85 sec, while the duration of an eclipse of this system is $\approx$300 sec (starting from the stream ingress), and its orbital period is 101 min \citep{2005v379}. The probability that at least 20\% of that 85 sec observation took place during eclipse is $\sim$6\%, which is non-negligible. The second possibility is that it may have been observed in a period of low accretion state, as is commonly seen in magnetic CVs \citep{2019bernardini}. Flickering and state changes have been observed in V379 Tel  \citep{2022ilkiewicz}, making it a possibility that it was in a low accretion state, with $L_X<9.8\times 10^{28} \textrm{erg s}^{-1}$ during the eRASS1 observation. Finally, in both systems, it could be possible that they are being viewed at an inclination such that the X-ray emitting region is obscured (by the disk in Gaia J154008.28–392917 or by the accretion stream in V379 Tel), leading to the absorption of X-ray photons \citep[e.g.][]{2017mukai}.

Importantly, we note that the non-detection of these two systems does not pose a challenge to our estimation of space densities, as the total number of systems is well within our 1$\sigma$ Poisson counting uncertainties. However, their non-detection does raise the possibility that a small fraction of CVs could be emitting at $L_X < 10^{29} \textrm{erg s}^{-1}$. Future eRASS releases will shine a light on these objects and reveal the cause of their current non-detection.

\subsection{Comparison to Theoretical Predictions and Previous Estimates of Space Densities}
\label{sec:prev_studies_densities}

\subsubsection{All CVs}
All population synthesis results predict an overabundance of CVs compared to observations. It has been postulated that a significant amount of optically faint, low $L_X$, CVs have been missed by previous observational campaigns \citep[e.g.][]{2012pretorius}, but our work rules that out. Our midplane number density is $\rho_{N,0} \approx 3.7 \pm 0.7\times 10^{-6} \textrm{ pc}^{-3}$, yet modern population synthesis analyses predict number densities of $\rho_N\approx 9\times 10^{-5} \textrm{ pc}^{-3}$ \citep{2018belloni}, assuming a classical canonical angular momentum loss (CAML) prescription, around 20 times higher than our inferred value. Adjusting the AML prescription to ``empirical" CAML, the predicted value drops to $\approx 2^{+2}_{-1}\times 10^{-5} \textrm{ pc}^{-3}$, which is still 5 times higher than our observed value \citep{2018belloni, 2020belloni} and in  $\gtrsim 2\sigma$ tension with our results. \cite{2015goliasch} predict a number density of $1.0 \pm 0.5 \times 10^{-5}  \textrm{ pc}^{-3}$, which is also in $\gtrsim 2\sigma$ tension with our results.

Previously, number densities of CVs were obtained in three different ways: 1) X-ray flux limited surveys, 2) large-scale optical spectroscopic surveys, and 3) the volume-limited sample of \cite{2020pala}, which assembled all known CVs out to 150 pc and based their completeness on SDSS targeting and CRTS observations.

We obtain more precise estimates than those carried out by previous X-ray surveys since they suffer from incompleteness. \cite{2012pretorius} predicted a number density of non-magnetic CVs of $\rho_N = 4^{+6}_{-2}\times 10^{-6} \textrm{ pc}^{-3}$, and \cite{2013pretorius} a number density of magnetic CVs of $8^{+4}_{-2}\times 10^{-7}\textrm{ pc}^{-3}$. Both analyses were carried out using data from the ROSAT catalogs, but an estimate of completeness could not be carried out due to a nonuniform selection and the challenge of isolating ROSAT sources at low Galactic latitudes ($|b| \lesssim 20^\circ$). Later, \cite{2018schwope} revisited systems from these samples, and obtained distances using newly available \textit{Gaia} parallaxes. \cite{2018schwope} obtained number densities ranging between $0.96 - 10.9 \times 10^{-6} \textrm{ pc}^{-3}$, depending on the particular sample used and the assumption of CV scale height, though individual estimates obtained similarly small error bars as the ones we show here (e.g. the use of the RASS sample with a 200 pc scale height yielded a number density of $4.7^{+0.7}_{-0.4}\times 10^{-6} \textrm{ pc}^{-3}$. 

We obtain similarly precise estimates than those carried out by optical spectroscopic surveys, yet those surveys suffer from incompleteness, different than that of X-ray surveys. \cite{2021inight} calculated a midplane number density of $\rho_0 = 2.2^{+1.0}_{-0.6} \times 10^{-6} \textrm{ pc}^{-3}$, assuming a scale height of 280 pc. They revisited their calculation in \cite{2023inight}, with new systems, and estimated a completeness factor of 2.14 which was used to compute a number density of  $\rho_0 = 7.23 \times 10^{-6} \textrm{ pc}^{-3}$ assuming a scale height of 260 pc (this means that before incorporating their completeness factor, they obtained a value of $\rho_0 = 3.37 \times 10^{-6} \textrm{ pc}^{-3}$). This is similar to the work of \cite{2020pala}, where a value of $\rho_0 = 3.7^{+0.6}_{-0.8} \times 10^{-6} \textrm{ pc}^{-3}$ was obtained, and then a completeness factor of 1.4 was incorporated to obtain a number density of $\rho_0 = 4.8^{+0.6}_{-0.8} \times 10^{-6} \textrm{ pc}^{-3}$, all assuming a scale height of 280 pc.  

All aforementioned X-ray campaigns have obtained number densities similar to ours and remarkably, the number densities derived by spectroscopic campaigns best agree with ours \textit{before} their application of a correction factor for completeness. The latter possibly suggests that such a correction has been overapplied. All in all, our results echo the conclusion of \cite{2023inight}, where it appears that various CV observational campaigns using independent methods are converging on a midplane number density of $\rho_{N,0} \approx 4-5\times 10^{-6} \textrm{ pc}^{-3}$. However, our work is unique in not needing to employ a completeness factor and appearing to be unbiased towards any CV subtype. 

\subsubsection{Magnetic CVs}

We calculate the midplane number density of magnetic CVs to be $\rho_\textrm{N, magnetic} = (1.3\pm0.5) \times 10^{-6 }\textrm{ pc}^{-3}$. The only theoretical estimate of magnetic CV space densities has been carried out by \cite{2020belloni}, where it was predicted that the space density of polars should be $5^{+5}_{-3} \times 10^{-6}  \textrm{ pc}^{-3}$ ($1.5^{+1.5}_{-0.6} \times 10^{-6}  \textrm{ pc}^{-3}$ excluding period bouncers), though the authors initially fix the fraction of polars relative to the entire CV population at 28 percent. Our number density is technically in disagreement with that, since our sample includes one IP (not considered by that analysis) and two magnetic period bouncers. 

It is most meaningful to compare our number density to that derived by \cite{2013pretorius}, $8^{+4}_{-2}\times 10^{-7}\textrm{ pc}^{-3}$, which agrees with ours within $1\sigma$. However, that sample is rich with more IPs and lacks period bouncers. Further analysis of our 300 pc and 1000 pc samples will reveal larger numbers of magnetic CVs and show at what distance the classic, high $L_X$ IPs will start to appear ($\approx$500 pc is suggested by \cite{2019suleimanov}).

\subsubsection{Period Bouncers}
We calculate the midplane number density of period bouncers to be $\rho_\textrm{N, PB} = 0.9 \pm 0.4 \times 10^{-6} \textrm{ pc}^{-3}$. All population synthesis results predict an overabundance of period bouncers compared to observations. Our work is the observational campaign that has found the largest space density of period bouncers to date, thanks to our sensitivity to low $L_X$ systems, where period bouncers live (Table \ref{tab:all}). Early works by \cite{1993kolb} predicted that $\approx70$ \% of CVs should be period bouncers, whereas modern works accommodate a range as low as 
$\approx$ 38\% \citep{2015goliasch} to $\approx$ 82\%. \citep{2020belloni}. Applying these percentages to global space densities,  \cite{2015goliasch} place the number density of period bouncers to be $\rho_\textrm{N,PB} \approx 4 - 7 \times 10^{-6} \textrm{ pc}^{-3}$ (at least 4 times higher than our value), and \cite{2020belloni} estimate it to be $16 \times 10^{-6} \textrm{ pc}^{-3}$ (nearly 20 times higher than our value), assuming the eCAML prescription. 

Only a handful of observational campaigns have placed meaningful constraints on the space density of period bouncers: 1) \cite{2018hernandezsanti} searched for eclipsing systems in the Palomar Transient Factory data archive and quantified their completeness using SDSS targeting to obtain an upper limit of $\rho_\textrm{N, PB} \lesssim 20\times 10^{-6} \textrm{ pc}^{-3}$, 2) \cite{2020pala} estimated the fraction of period bouncers in their sample to be 7--14\%, yielding a number density of $\approx 0.3-0.6 \times 10^{-6} \textrm{ pc}^{-3}$, and 3) \cite{2023inight} analyzed SDSS I--V data to estimate a value of $\approx0.2\times10^{-6} \textrm{ pc}^{-3}$. 

In contrast to virtually all previous studies, we obtain a value of $\rho_\textrm{N, PB} = 0.9 \pm 0.4 \times 10^{-6} \textrm{ pc}^{-3}$. This high value is due to the inclusion of two magnetic period bouncers not mentioned in the \cite{2020pala} sample: 1) the reclassification of SDSS J125044.42+154957.3 by \cite{2023munozgiraldo} as a magnetic period bouncer (this was not the case in \cite{2020pala}), and 2) our discovery of 1eRASS J054726.9+132649. This suggests that other nearby magnetic period bouncers may be lurking, and could be revealed by X-ray surveys from eROSITA \citep[e.g.][]{2024galiullin}. However, while previous observational campaigns have likely underestimated the space density of period bouncers, the number we obtain is still at least 4 times less than that predicted by the most conservative population synthesis estimates. Our work is among the most conclusive to date that shows the need for a modification to the standard CV evolutionary picture --- for example, either through a modification of the magnetic braking prescription \citep{2024sarkar} or through the detachment of CVs as they pass through the period minimum \citep{2023schreiber}. 

\subsubsection{AM CVns}
We calculate the midplane numebr density of AM CVns to be $\rho_\textrm{N, AM CVn} = 5.5 \pm 3.7 \times 10^{-7} \textrm{ pc}^{-3}$ We are able to place meaningful constraints on the space density of AM CVns because of our straightforward selection function. Even with three systems in our sample, our 68\% Poisson intervals place the errors of our number density estimates on par with those of other observational campaigns. Only a handful of studies have estimated the space density of AM CVns due to 1) the intrinsic rarity of systems and 2) the difficulty in setting up a systematic survey to identify long-period, low accretion rate systems. Because, like CVs, all AM CVns are X-ray emitters \citep[e.g.][]{2005ramsay, 2006ramsay, 2023begari}, it is difficult for nearby systems to be missed with the X-ray Main Sequence. 

Early work on population synthesis of AM CVns predicted space densities of $\rho_\textrm{N, AM CVn} = 0.4-1.7\times10^{-4} \textrm{ pc}^{-3}$ \citep{2001nelemans}, at least 100 times higher than our derived value. This incredibly high number would mean that AM CVns would have higher space densities than CVs. 

Two types of observational campaigns have placed meaningful constraints on the space density of AM CVns, with the first being spectroscopic surveys from SDSS. \cite{2007roelofs} and \cite{2013carter} calculated number densities of $1-3\times10^{-7} \textrm{ pc}^{-3}$ and $5\pm3 \times 10^{-7} \textrm{ pc}^{-3}$, respectively. The other method has been recently carried out by \cite{2022vanroestel}, who systematically searched the ZTF database for eclipsing AM CVns and derived a space density of $6^{+6}_{-2}\times10^{-7} \textrm{ pc}^{-3}$. That study effectively ruled out a large population of faint, very low accretion rate systems, meaning that our X-ray survey will not miss a large population of AM CVns. It is worth noting that \cite{2018ramsay} also derived the space density of AM CVns from the collection of all systems known up until that point. There, it was shown that the $1\sigma$ lower limit of the AM CVn number density was $>3\times10^{-7} \textrm{ pc}^{-3}$. 

Our midplane number density is consistent within $1\sigma$ with values derived from both spectroscopic surveys and searches for eclipsing systems. However, it is lower than the expected number density of \cite{2001nelemans} by at least two orders of magnitude.  Our result reinforces the tension with AM CVn population synthesis estimates, and demonstrates the need for a modern recalculation. This is especially important for estimating the population resolved by LISA and that which will make up the background \citep{2004nelemans, 2024kupfer, 2024lisa}.

\section{Conclusion}
\label{sec:conclusion}

We have conducted a systematic, volume-limited survey of CVs down to the lowest X-ray luminosity to date, $L_X=1.3\times 10^{29} \textrm{ erg s}^{-1}$ in the 0.2--2.3 keV range. Our main survey extends to a distance of 150 pc, and was possible thanks to a crossmatch of the SRG/eROSITA eRASS1 X-ray survey of the Western Galactic Hemisphere and \textit{Gaia} DR3. We employed a novel tool called the ``X-ray Main Sequence" --- a color-color diagram with \textit{Gaia} BP--RP on the horizontal axis and $F_X/F_\textrm{opt}$ on the vertical axis --- to create our sample. We recover virtually all previously known CVs within 150 pc in this part of the sky and have discovered two new systems. Our main conclusions are:
\begin{enumerate}
    \item The X-ray Main Sequence is a highly efficient tool for systematic searches of the X-ray + optical sky (Figure \ref{fig:all_samples}). Within 150 pc, we selected 28 CV candidates, a mere $\approx$0.1\% of the SRG/eROSITA eRASS1 + \textit{Gaia} parent sample (Table \ref{tab:volume_samples}). We discarded three false matches, two non-CV compact object binaries, and were left with 23 CVs within 150 pc in the Western Galactic Hemisphere.
    \item We discovered two new CVs in the 23 within 150 pc: 1eRASS J054726.9+132649, a magnetic period bouncer (Figure \ref{fig:pb}), and 1eRASS J101328.7-202848, a long-period AM CVn which is now the third closest ultracompact system to Earth, at 140 pc (Figure \ref{fig:amcvn}). 
    \item Using our 150 pc sample, we constructed luminosity functions and showed a ``flattening" at $L_X\sim 10^{30} \textrm{erg s}^{-1}$ for the first time in CVs (Figure \ref{fig:xlf}). We calculated the CV number density at the midplane to be $\rho_\textrm{N,CV} = 3.7\pm 0.7\times 10^{-6} \textrm{ pc}^{-3}$ (Table \ref{tab:xlf}). This number density agrees to within 1$\sigma$ with most previous estimates from spectroscopic surveys (without completeness correction) and X-ray flux limited surveys. However, it is still lower than that predicted by theory by a factor of $\sim$5--20. 
    \item We calculated the AM CVn number density at the midplane to be $\rho_\textrm{N,AM CVn} = 5.5\pm 3.7\times 10^{-7} \textrm{ pc}^{-3}$ (Table \ref{tab:xlf}). Our novel selection method allows us to be complete, even while only having a few systems. This agrees to within 1$\sigma$ with most previous estimates from spectroscopic and optical photometric surveys. However, it is still lower than that predicted by theory by a factor of $\sim$70--200. 
    \item The average X-ray luminosity of CVs in our 150 pc sample is $\langle L_X \rangle \sim 10^{30} \textrm{erg s}^{-1}$ in the 0.2--2.3 keV range (Figure \ref{fig:lum_dist}). In contrast, $L_X$ distributions built from X-ray matches to primarily optically-discovered CVs yield $\langle L_X \rangle \sim 10^{31}-10^{32} \textrm{erg s}^{-1}$, which means that $\langle L_X \rangle$ of CVs has been systematically overrestimated in the past.
    \item The fraction of magnetic CVs in our 150 pc sample is $35\%$, and that of period bouncers is $25\%$, assuming Poisson counting errors in the main sample. This fraction of period bouncers is higher than all previous observational campaigns, but still lower than the 40--70\% predicted by population synthesis.
\end{enumerate}

By focusing on the volume out to 150 pc from the Sun, this work is somewhat biased to the low-$L_X$ end of CVs. Nevertheless, our local survey should be representative of other regions in the Milky Way that resemble the solar neighborhood. Furthermore, low-$L_X$ systems such as AM CVns and period bouncers have been the most difficult to find in previous surveys, as highlighted by the discovery of two new systems in our sample. It has only been with recent surveys made possible by SRG/eROSITA that such systems have come to light \citep[e.g.][]{2023rodriguez_amcvn, 2024galiullin}. In contrast, high-$L_X$ systems like bright IPs have been easier to discover in the past, but are more rare, making them more distant and not recovered by our search \citep[e.g.][]{2019suleimanov}.  

Future work should spectroscopically verify CV candidates in our 300 pc and 1000 pc samples, which we make publicly available. Since the number of systems approaches the hundreds, this could be best accomplished using large spectroscopic surveys such as SDSS-V \citep{2017sdssv, 2023sdssv}, the Dark Energy Spectroscopic Instrument \citep[DESI;][]{2016desi}, the 4-metre Multi-Object Spectrograph Telescope \citep[4MOST;][]{20194most}, and the William Herschel Telescope Enhanced Area Velocity Explorer \citep[WEAVE;][]{2012weave}. 

Upcoming eRASS releases will probe the X-ray variability of these objects, and trends regarding X-ray hardness ratios of CVs and AM CVns should be investigated on a large scale. Co-added eROSITA surveys will go a few times deeper than eRASS1, and will discover many new systems, particularly at larger distances. For example, \cite{2024schwope} recently compiled a list of the compact white dwarf binaries in the eFEDS survey of SRG/eROSITA, covering 140 deg$^2$ of the sky. In that work, of the 23 systems, only 9 systems would pass the flux limit of eRASS1. The remaining systems are at $\gtrsim250$ pc, suggesting that our 300 pc sample, and in particular our 1000 pc sample would greatly benefit from deeper eRASS surveys as Figure \ref{fig:lum_dist} demonstrates. At the eRASS:4/eRASS:8 limit of $F_X = 1\times 10^{-14}/7\times 10^{-15}$ $\textrm{erg s}^{-1}\textrm{cm}^{-2}$ \citep[e.g.][]{2021sunyaev}, a survey complete down to $L_X = 10^{29} \textrm{erg s}^{-1}$ can extend out to 300 pc / 360 pc.

New optical surveys will be useful in the near future as well. Releases with updated astrometry, photometry, and ultra low-resolution spectroscopy from \textit{Gaia} DR4/DR5 will also add to the complete picture of CVs and AM CVns. These releases will improve parallaxes to faint, WD-dominated systems at $\gtrsim300-400$ pc, which are currently difficult to detect due to a combination of large parallax errors and faintness at that distance \citep[e.g.][]{2021gentile}. Also on the horizon is the Rubin Observatory Legacy Survey of Space and Time (LSST), which will use large, optically selected samples to provide further data on faint systems, particularly AM CVns and eclipsing and outbursting CVs in the period bounce regime \citep[e.g.][]{2015lsst}. In both cases, combining SRG/eROSITA data and employing the X-ray Main Sequence will be a useful tool in identifying candidate compact object binaries and forming large, uniformly selected samples as we have done here.

\section{Acknowledgements}
We thank Roman Krivonos for insightful feedback, Kevin Burdge, Dovi Poznanski, and Jim Fuller for useful discussions, and Sunny Wong for providing AM CVn evolutionary models. ACR acknowledges support from an NSF Graduate Fellowship. 

ACR thanks the LSST-DA Data Science Fellowship Program, which is funded by LSST-DA, the Brinson Foundation, and the Moore Foundation; his participation in the program has benefited this work. RLO is a Research Fellow of the Brazilian institution CNPq (PQ-315632/2023-2).

This work is based on data from eROSITA, the soft X-ray instrument aboard SRG, a joint Russian-German science mission supported by the Russian Space Agency (Roskosmos), in the interests of the Russian Academy of Sciences represented by its Space Research Institute (IKI), and the Deutsches Zentrum f{\"u}r Luft- und Raumfahrt (DLR). The SRG spacecraft was built by Lavochkin Association (NPOL) and its subcontractors, and is operated by NPOL with support from the Max Planck Institute for Extraterrestrial Physics (MPE). The development and construction of the eROSITA X-ray instrument was led by MPE, with contributions from the Dr. Karl Remeis Observatory Bamberg \& ECAP (FAU Erlangen-Nuernberg), the University of Hamburg Observatory, the Leibniz Institute for Astrophysics Potsdam (AIP), and the Institute for Astronomy and Astrophysics of the University of T{\"u}bingen, with the support of DLR and the Max Planck Society. The Argelander Institute for Astronomy of the University of Bonn and the Ludwig Maximilians Universit{\"a}t Munich also participated in the science preparation for eROSITA.

This work presents results from the European Space Agency (ESA) space mission Gaia. Gaia data are being processed by the Gaia Data Processing and Analysis Consortium (DPAC). Funding for the DPAC is provided by national institutions, in particular the institutions participating in the Gaia MultiLateral Agreement (MLA). The Gaia mission website is \url{https://www.cosmos.esa.int/gaia}. The Gaia archive website is \url{https://archives.esac.esa.int/gaia}.

Some of the data presented herein were obtained at Keck Observatory, which is a private 501(c)3 non-profit organization operated as a scientific partnership among the California Institute of Technology, the University of California, and the National Aeronautics and Space Administration. The Observatory was made possible by the generous financial support of the W. M. Keck Foundation. 
The authors wish to recognize and acknowledge the very significant cultural role and reverence that the summit of Maunakea has always had within the Native Hawaiian community. We are most fortunate to have the opportunity to conduct observations from this mountain. We are also grateful to the staff of Palomar Observatory and that of Lick Observatory for their assistance in carrying out observations used in this work.

Based on observations obtained with the Samuel Oschin Telescope 48-inch and the 60-inch Telescope at the Palomar Observatory as part of the Zwicky Transient Facility project. ZTF is supported by the National Science Foundation under Grants No. AST-1440341 and AST-2034437 and a collaboration including current partners Caltech, IPAC, the Weizmann Institute of Science, the Oskar Klein Center at Stockholm University, the University of Maryland, Deutsches Elektronen-Synchrotron and Humboldt University, the TANGO Consortium of Taiwan, the University of Wisconsin at Milwaukee, Trinity College Dublin, Lawrence Livermore National Laboratories, IN2P3, University of Warwick, Ruhr University Bochum, Northwestern University and former partners the University of Washington, Los Alamos National Laboratories, and Lawrence Berkeley National Laboratories. Operations are conducted by COO, IPAC, and UW.

Software used: Python and the following libraries: \texttt{matplotlib} \citep{matplotlib}, \texttt{scipy} \citep{scipy}, \texttt{astropy} \citep{astropy}, \texttt{numpy} \citep{numpy}. \texttt{PypeIt} \citep{2020pypeit}, \texttt{lpipe} \citep{2019perley_lpipe}, and Tool for OPerations on Catalogues And Tables \texttt{(TOPCAT)} \citep{topcat}.

\bibliography{main}{}
\bibliographystyle{aasjournal}

\appendix
\section{CV Selection with the X-ray Main Sequence}
\label{sec:selection}
\begin{figure*}
    \centering
    \includegraphics[width=\textwidth]{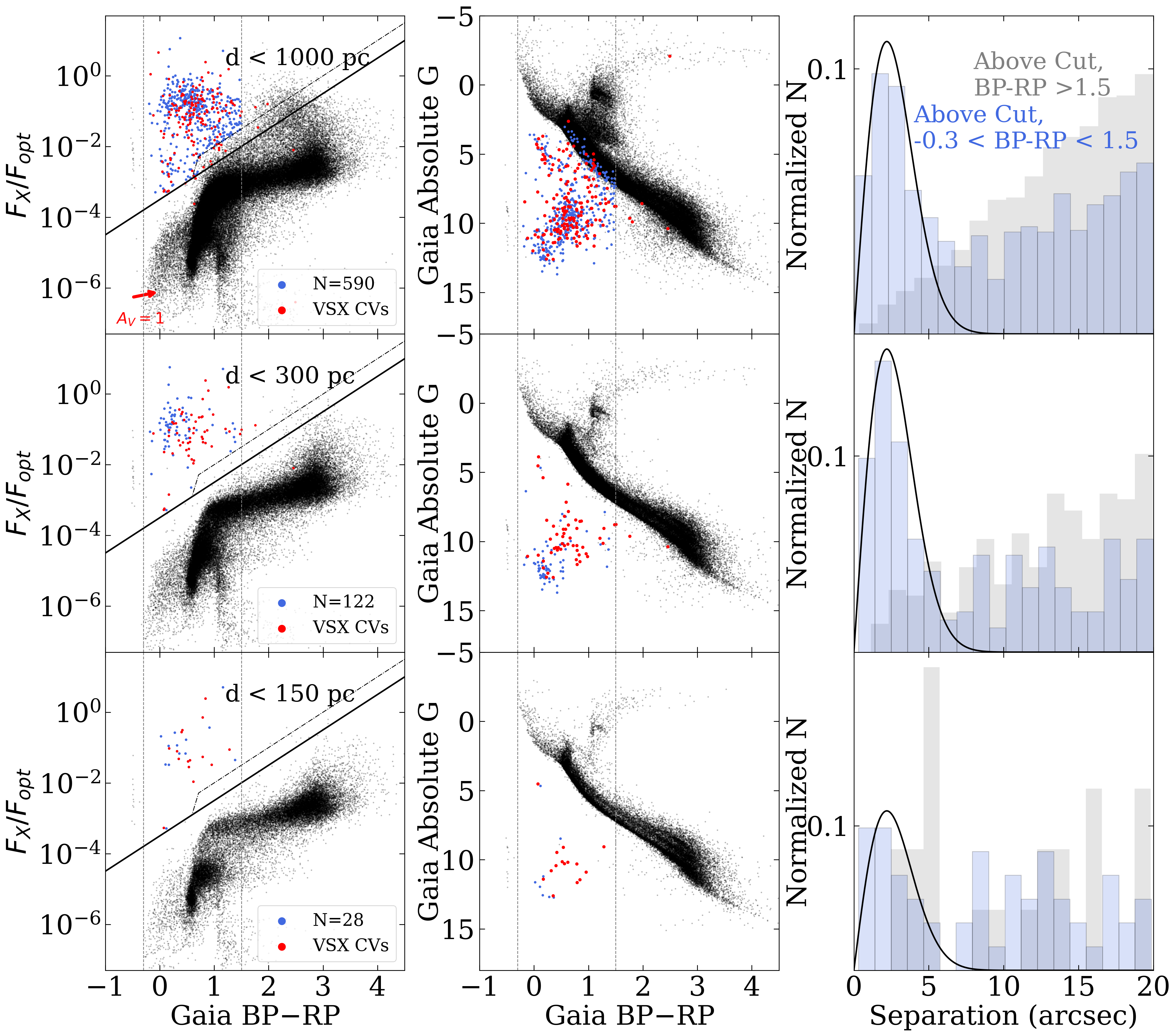}
    \caption{Our volume-limited samples are shown on the X-ray Main Sequence (left) and HR diagram (center), with black points denoting all objects in the sample, blue our CV candidates, and red the best vetted CVs from the VSX catalog. Only a few VSX CVs are below the ``modified" cut (dash dot line) on the left, meaning that this can be used in place of the cut from \cite{r24} (solid diagonal line). On the right are distributions of the separation between X-ray and optical points. A Rayleigh distribution with $\sigma_\textrm{sep}=2.2"$ (black line) best fits the final sample at all distances, justifying our choice of cuts.}
    \label{fig:selection_justify}
\end{figure*}

Here, we justify our choice of enforcing the -0.3 $<$ BP--RP $<$ 1.5 cut , as well as our ``modified" cut in the X-ray Main Sequence when creating our samples. At all distances, we crossmatched our volume limited sample with a clean set of CVs from the VSX catalog (see Section \ref{sec:ref_catalog}), where we enforced that the orbital period also be well known (no ``:" in the period category) to ensure the best selection. We present the results in Figure \ref{fig:selection_justify}. 

In Figure \ref{fig:selection_justify}, it is clear that none of the VSX CVs fall under the modified cut (dash dot line) in the 150 pc and 300 pc samples. Less than 3\% of systems fall below the cut in the 1000 pc sample, with $<1\%$ falling below the empirical cut from \cite{r24} (solid diagonal line). No CVs are located in the BP--RP $<-0.3$ region at any distance. This means that the systems located there are likely hot WDs thermally emitting X-rays, though that should be investigated in future studies. 

Finally, no VSX CVs are located in the BP--RP $>1.5$ region in the 150 pc sample, while 2 ($<3\%$) are there in the 300 pc sample and 5 ($<1\%$) are there in the 1000 pc sample. It is especially clear in the X-ray Main Sequence of the 1000 pc sample (upper left panel of Figure \ref{fig:selection_justify}) that there is a significant amount of objects above the ``empirical cut" \citep{r24} as well as ``modified cut", redward of BP--RP of 1.5. The distribution of the separation between X-ray and optical points (right panels of Figure \ref{fig:selection_justify}) shows that the distribution of those points (shown in gray), however, does not follow a Rayleigh distribution and is instead consistent with randomly associated matches. In contrast, the distribution of the our CV candidates (shown in blue), clearly follows a Rayleigh distribution, with a best-fit $\sigma_\textrm{sep}$ (Equation \ref{eq:dist}) of 2.2" at all distances. We clarify that the blue distributions plotted on the right panels of Figure \ref{fig:selection_justify} are comprised of \textit{all} possible matches out to 20". The candidates we plot on the left and center panels (and comprise our samples), include only crossmatches out to 7.7" (3.5$\sigma_\textrm{sep}$; see Section \ref{sec:sample}), thus being dominated by true associations. This exercise shows the importance of 1) starting with a larger crossmatch radius than the nominal positional error of a survey, and 2)always plotting the distribution of separations when doing a crossmatch. A blind crossmatch, accepting points redward of BP--RP = 1.5 in this case, would have resulted in being dominated by false matches, \textit{even when only keeping points at very low separations} (e.g. $<5$ arcsec).

\section{CV Scale Height Calculation}
\label{sec:h_cv}

In order to be self consistent with the Galactic mass profile analysis carried out by \cite{2016barros}, we effectively assumed a CV scale height, $h_\textrm{CV}$ of 205 pc. Calculating $h_\textrm{CV}$ for CVs is difficult, and technically must be done for different CV subtypes separately to account for their being in different evolutionary stages and therefore having different intrinsic ages \citep[e.g.][]{2007pretorius}. However, it would likely require a much larger, spectroscopically confirmed sample of systems to undertake such an analysis. We proceed by computing a scale height of CVs for our largest samples, but explain why simply setting $h_\textrm{CV}$ to a fixed value is justified.

Because the scale height of the thin disk is $\approx$200 pc, we cannot compute a scale height based on the 150 pc sample alone. However, we can make use of three other samples in this study: our 1000 pc sample, the VSX + eROSITA sample, and the VSX + ROSAT sample, which are sensitive to CVs far enough away to trace their characteristic $\textrm{h}_\textrm{CV}$. We exclude systems in the VSX samples beyond 1000 pc in order to have a fair comparison to our 1000 pc sample.

We follow a similar approach to \cite{2008Revnivtsev} and \cite{2022suleimanov}, and assume that CVs follow an exponential distribution in the Milky Way, with the probability of a system to be located at a height $z$ above the Galactic place to be:
\begin{gather}
    P_i(z_i) = \frac{1}{h_\textrm{CV}}\exp\left(\frac{-|z_i|}{h_\textrm{CV}}\right)
\end{gather}
where $z$ is defined as in Equation \ref{eq:coords}. We can compute a log-likelihood function:
\begin{gather}
    \log \mathcal{L}(z) = \log \prod_i P_i (z_i) = -\sum_i^N \ln h_\textrm{CV}  - \frac{|z_i|}{h_\textrm{CV}}
\end{gather}

\begin{figure}
    \centering
    \includegraphics[width=0.5\linewidth]{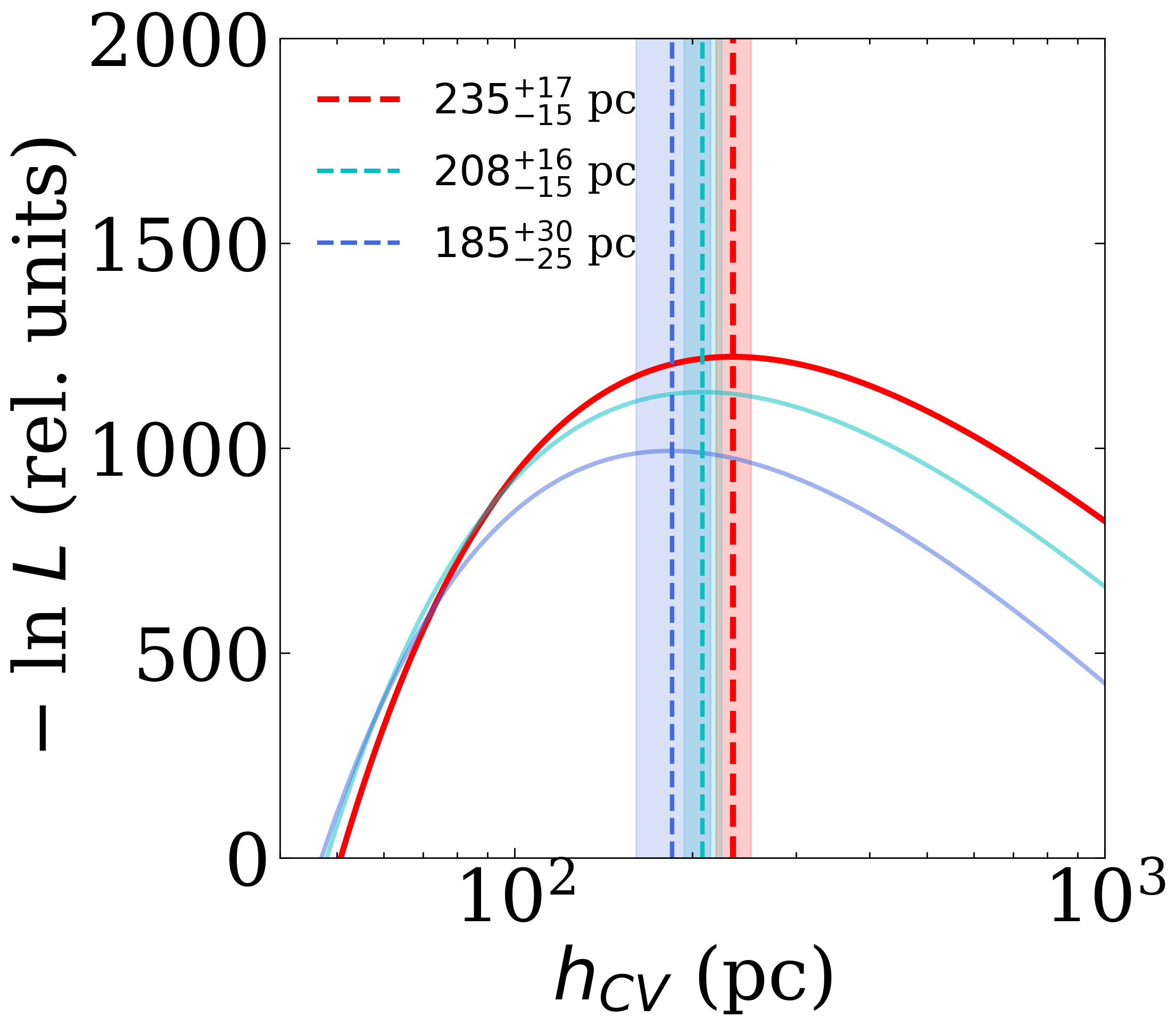}
    \caption{Negative log likelihood of the CV characteristic scale height, $h_\textrm{CV}$ for the VSX + ROSAT (blue), VSX + eROSITA (cyan), and our 1000 pc sample (red). We plot the most likely value as a dotted line and show 95\% confidence intervals as shaded regions. The ROSAT sample likely has a smaller value since it traces X-ray bright systems (e.g. polars, IPs) which are younger, while the VSX + eROSITA and our 1000 pc sample are more sensitive to X-ray faint systems farther along in their evolution. Both the VSX + eROSITA and our 1000 pc sample are consistent with our fixed value of $h_\textrm{CV} = 205$ pc.}
    \label{fig:h_cv}
\end{figure}

We compute the value that maximizes the negative log-likelihood and present that value along with 95\% confidence intervals in Figure \ref{fig:h_cv}. The ROSAT sample suggests that $h_\textrm{CV} \approx 185$ pc since it traces X-ray bright systems (e.g. polars, IPs) which are younger, while the VSX + eROSITA and our 1000 pc sample are more sensitive to X-ray faint systems farther along in their evolution, and instead suggest that $h_\textrm{CV} \approx 200-240$ pc. At the end of the day, neither of those two samples can be taken as ground truth, since the VSX + eROSITA sample suffers from incompleteness and our 1000 pc sample of CV candidates is not spectroscopically verified. 

As a final calculation, we determine the midplane number density of CVs assuming different scale heights from our 150 pc sample, assuming a Galactic model as in Equation \ref{eq:mass}. Assuming a extreme values of $h_\textrm{CV}=$100, 500 pc, respectively, we obtain midplane number densities of $\rho_{N,0} \approx 4.7\times10^{-6} \textrm{ pc}^{-3}$ and $3.2\times10^{-6} \textrm{ pc}^{-3}$, respectively, which lead to no more than $\approx$ 30\% deviations from our calculated values in Table \ref{tab:xlf}.

\end{document}
